\pdfoutput=1
\documentclass[12pt]{elsarticle}
\usepackage{blindtext}
\usepackage{amsmath}

\usepackage[utf8x]{inputenc}
\usepackage[T1]{fontenc}
\usepackage{tikz}
\usepackage{pgfplots, pgfplotstable}
\usetikzlibrary{calc,arrows,fadings,decorations.pathreplacing,decorations.markings,patterns,shapes.geometric}
\usepackage{fp}
\usepackage{yhmath}
\usepackage{verbatim}

\usepackage{color,soul}

\usepackage{amssymb}
\usepackage{caption}
\usepackage{subcaption}
\usetikzlibrary{calc,fit,intersections,shapes}
\usetikzlibrary{calc}
\usepgfplotslibrary{polar}
\pgfplotsset{compat=1.6}
\pgfplotsset{every axis plot}
\journal{JMPS}

\usepackage{amsmath}
\usepackage[normalem]{ulem}
\usepackage{multirow}

\title{Design of Piezoelectric Metastructures with Multi-Patch Isogeometric Analysis for Enhanced Energy Harvesting and Vibration Suppression}

\author[label1]{P. Peralta-Braz}
\author[label1]{M. M. Alamdari}
\author[label2]{M. Hassan}
\author[label1]{E. Atroshchenko\footnote{Corresponding author, e.atroshchenko@unsw.edu.au, eatroshch@gmail.com}}

\address[label1]{School of Civil and Environmental Engineering, University of New South Wales, Sydney, Australia}
\address[label2]{School of Computer Science and Engineering, University of New South Wales, Sydney, Australia}

\begin{document}

\begin{abstract}

Metastructures are engineered systems composed of periodic arrays of identical components, called resonators, designed to achieve specific dynamic effects, such as creating a bandgap-a frequency range where waves cannot propagate through the structure. When equipped with patches of piezoelectric material, these metastructures exhibit an additional capability: they can harvest energy effectively even from frequencies much lower than the fundamental frequency of an individual resonator. This energy harvesting capability is particularly valuable for applications where low-frequency vibrations dominate. To support the design of metastructures for dual purposes, such as energy harvesting and vibration suppression (reducing unwanted oscillations in the structure), we develop a multi-patch isogeometric model of a piezoelectric energy harvester. This model is based on a piezoelectric Kirchhoff-Love plate—a thin, flexible structure with embedded piezoelectric patches—and uses Nitsche's method to enforce compatibility conditions in terms of displacement, rotations, shear force, and bending moments across the boundaries of different patches. The model is validated against experimental and numerical data from the literature. We then present a novel, parameterized metastructure plate design and conduct a parametric study to explore how resonator geometries affect key performance metrics, including the location and width of the band gap and the position of the first peak in the voltage frequency response function. This model can be integrated with optimization algorithms to maximize outcomes such as energy harvesting efficiency or vibration reduction, depending on application needs.

\end{abstract}

\maketitle

\section{Introduction}
\label{S:1}
Mechanical metamaterials are artificial structures characterised by their distinctive mechanical attributes determined by structural configurations rather than their material composition \cite{lee2022piezoelectric}. Typically, their structures follow a periodic pattern, aiming to achieve unconventional, adaptable, and naturally non-existent properties \cite{han2023origami}. Mechanical metamaterials can exhibit a variety of unique properties, including bandgap phenomena \cite{wei2021smp,sugino2017general}, heat manipulators \cite{jansari2022design}, enhanced deformation resistance \cite{jiao2020hierarchical}, negative Poisson’s ratio \cite{mizzi2018mechanical}, energy absorption \cite{zhang2022dual}. These properties are achieved through adequately designing their substructure unit cells \cite{xiong2023optimization}. 

In recent years, there has been extensive study of mechanical metamaterial-inspired structures, or simply metastructures, which consist of periodic cell arrays featuring internal resonators. Metastructures are effective in creating locally resonant bandgaps that prevent the propagation of vibration waves within specific frequency ranges through the elastic medium \cite{lee2022piezoelectric}. The formation of a bandgap is attributed to the resonance of internal unit cells, which causes a significant alteration in the effective properties of the metastructure near the resonant frequency of the unit cell \cite{khattak2022concurrent}. Its phenomenon is particularly useful in the fields of vibration isolation, soundproofing, and seismic mitigation.

Metastructures have been studied for suppressing vibrations in engineering structures such as beams. Typically, the internal resonators are modelled as masses connected to the main structure via springs, demonstrating the formation of an attenuation bandgap through both theorical and experimental studies \cite{xiao2012broadband, shuguang2017studies, sun2010theory, sugino2016mechanism, sachdeva2024aperiodicity, wang2023enhancement, patro2023vibration}. Some researches have explored the use of graded metastructures, which involve variations in the natural frequency of each resonator \cite{hu2018internally,hu2021metamaterial}. The implementation of graded local resonators has resulted in an increased attenuation bandgap compared to cases with non-graded local resonators \cite{hu2021metamaterial}.

Building on the previous findings, researchers have been exploring the multifunctional applications of metastructures that leverage local-resonance mechanisms in bandgap formation, offering a promising solution for both vibration mitigation and low-power energy harvesting \cite{sugino2018analysis,sugino2018merging,zhao2022graded}. These configurations can effectively suppress vibrations while simultaneously generating electricity from ambient vibrations, enabling the powering of low-power devices such as wireless sensors and wearable technology \cite{covaci2020piezoelectric,sezer2021comprehensive}. The integration of metastructures within the Internet of Things aligns with the goals of the Fourth Industrial Revolution, opening innovative applications in various fields \cite{lee2022piezoelectric, izadgoshasb2021piezoelectric}.

The energy harvesting mechanism utilises the direct piezoelectric effect, whereby mechanical stress applied to a piezoelectric material generates an electrical voltage \cite{erturk2011piezoelectric,erturk2008distributed}. To exploit this effect, piezoelectric material sheets are strategically incorporated into the local resonators of the metastructures. In general, an array of parasitic beam-like resonators are strategically covered with piezoelectric materials, facilitating efficient energy harvesting. While graded local resonators have been explored \cite{zhao2022graded, el2020multiple}, recent studies have innovated by using the mechanism inspired on acoustic–elastic metamaterial \cite{hu2017metastructure} or, moving away from local resonators, employing peanut-shaped periodic auxetic structures completely covered with piezoelectric material \cite{zhang2024numerical}. Therefore, the integration of piezoelectric material and metastructures has yielded superior performance for energy harvesting, outperforming conventional configurations \cite{lee2022piezoelectric,khattak2022concurrent}. Specifically, this enhancement extends to improved operation across multiple frequencies, with notable effectiveness in energy harvesting at low frequencies \cite{zhao2022graded}.

Piezoelectric sheets can be used not only for energy harvesting but also in bandgap formation. An array of piezoelectric patches, coupled with resonant shunt circuits (e.g., resistive and inductive elements), can be integrated into a host structure, similar to an array of purely mechanical resonator units. This integration avoids the propagation of vibration waves through the elastic medium, ensuring bandgap formation \cite{sugino2018analysis,aghakhani2020modal,dwivedi2021optimal,mao2024analytical}. This approach does not add significant mass or volumetric occupancy to the system. In particular, the use of graded resonators has been widely studied, demonstrating significant improvements in bandgap creation \cite{hu2021metamaterial, jian2022design,alshaqaq2020graded,liu2023broadband,jian2023analytical, schimidt2023piezoelectric, jian2022adaptive}.

The design process of the unit cells is crucial to achieving optimal performance. In this regard, a fast, accurate, and versatile numerical model has become a fundamental tool in the design process. The electromechanically coupled finite element plate model \cite{junior2009electromechanical} discretised with isogeometric analysis (IGA) \cite{peralta2020parametric} arises as a good option. The model is based on the Kirchhoff-Love plate theory and Hamilton’s principle. IGA can accurately represent second-order derivatives in the Kirchhoff-Love plate model due to the C$^1$ or higher-order continuous Non-Uniform Rational B-Splines (NURBS) basis functions leading to an overall higher accuracy per Degree of Freedom (DOF) compared to the standard Finite Element Method (FEM) \cite{nguyen2017geometrically,hughes2005isogeometric}. 

Moreover, IGA is capable of accurately representing complex shapes, making it versatile in a wide range of configurations. However, the number of configurations that can be represented using single-patch NURBS models is limited. As geometric complexity increases, the number of patches required to represent a structure also increases. In multi-patch analysis, the different patches involved must be connected along their patch interfaces \cite{schuss2019multi}. A strong coupling based on degrees of freedom belonging to the control points of adjacent patches generally ensures only C$^0$-continuity across the coupled domains, considering that the Kirchhoff–Love plate requires at least a C$^1$-continuous patch \cite{guo2015nitsche}. 

The continuity issues that arise from using a strong coupling approach have inspired the development of weak boundary condition enforcement strategies. These strategies can be used to enforce both displacement and normal rotation boundary conditions. One such method is Nitsche’s method, which is variationally consistent and does not introduce additional degrees of freedom that need to be solved \cite{schuss2019multi, guo2015nitsche, benzaken2021nitsche,  guo2015weak}. The coercivity of the governing problem can be ensured with additional stability terms without compromising the conditioning properties of the algebraic equations \cite{guo2015nitsche}.
 
The paper's contribution focuses on the explicit formulation of a multi-patches piezoelectric plate devices model based on the Kirchhoff–Love plate. The framework incorporates Nitsche’s method formulation to effectively enforce weakly essential boundary conditions on the patch interface, which is explicitly deducted for piezoelectric harvester configurations. To validate the proposed model, various benchmark problems from the existing literature are employed. Furthermore, the model is applied to investigate metamaterial piezoelectric cases, providing an in-depth analysis of its performance. This analysis encompasses both well-established cases from the literature and novel configurations introduced by the authors.

The paper is structured as follows. Section \ref{S:2} presents the theoretical framework of the model, which includes Nitsche-based formulations for the patch interface conditions. In Section \ref{S:3}, the model is explicitly formulated with NURBS-based Isogeometric Analysis (IGA) after discretization. The model is then validated through rigorous testing against various benchmark problems derived from existing literature in Section \ref{S:4}. In Section \ref{S:5}, the model is employed to investigate a metamaterial configuration based on mechanical resonators, and its findings are compared with experimental results available in the literature. Section \ref{S:6} explores the versatility of the model by investigating a metamaterial scenario that involves piezoelectric shunt damping. In Section \ref{S:7}, the authors delve into a novel metamaterial configuration. Conclusions are presented in Section \ref{S:8}.

\section{Formulation of the multi-patch PEH model}
\label{S:2}
\begin{figure}[ht]
	\centering
	\includegraphics[width=0.9\textwidth]{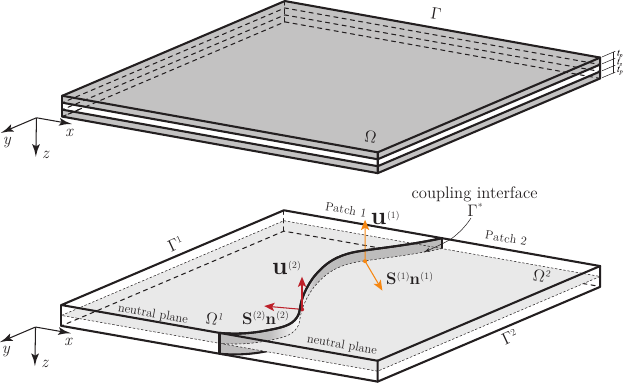}
\caption{Schematics of the weak coupling of plate patches in Kirchhoff-Love plate.}
	\label{Ch6:fig:schemaNit}
\end{figure}

A PEH is typically modelled as a plate, comprising a primary substructure with piezoelectric material bound to its upper and lower surfaces, this configuration is commonly referred to as a bimorph. Figure \ref{Ch6:fig:schemaNit}a illustrates an example where $\Omega$ represents the open-bounded region with boundary $\Gamma$. This boundary is the projection onto the $x-y$ plane for all layers of the PEH. These layers have a uniform thickness, with $t_s$ denoting the thickness of the substructural layer and $t_p$ representing the thickness of the piezoelectric layer with total thickness $t = t_s + 2t_p$. The domain of the PEH can be divided into several non-overlapping patches. Figure \ref{Ch6:fig:schemaNit}b illustrates the partitioning of the domain $\Omega$ into two distinct subdomain patches, denoted as $\Omega^{(m)}$. Here, $m$ signifies the specific patch, specifically $m$=1, 2 in this case. Furthermore, excluding the internal boundaries of the patches, represented as $\Gamma^*$, the boundary within each body is denoted as $\Gamma^{(m)}$. The normal vectors, denoted as $\boldsymbol{n}^{(m)} $, are defined on the internal boundary $\Gamma^*$ and point outward from their respective domains. For the purpose of formulating the model, when patches 1 and 2 are interconnected at the $\Gamma^*$ interface, the normal unit vector is considered to be $\boldsymbol{n} = \boldsymbol{n}^{(1)} = -\boldsymbol{n}^{(2)} = \{n_x, n_y, n_z \}$. The 3D domain of the substructure layer is encompassed by $V_s^{(m)} = \Omega \times [-\frac{t_s}{2}, \frac{t_s}{2}]$, whereas the two piezoelectric layers occupy the domain $V_p^{(m)} = \Omega \times ([-\frac{t}{2}, -\frac{t_s}{2}] \cup [\frac{t_s}{2}, \frac{t}{2}])$.

According to the Kirchhoff-Love plate theory, transverse shear strains are negligible, and the transverse normal remains perpendicular to the median plane. Consequently, the displacements $u^{(m)}$ and $v^{(m)}$ are assumed to result solely from the plate's bending cross-section rotation. As a result, the total displacement field $\boldsymbol{u}^{(m)}$ is defined as,
\begin{equation}
    \label{u}
   \boldsymbol{u}^{(m)}  = \left\{u^{(m)}, v^{(m)}, w^{(m)} \right\}^{T}  = \left\{-z\dfrac{\partial w^{(m)}}{\partial x}, -z\dfrac{\partial w^{(m)}}{\partial y}, w^{(m)} \right\}^{T}  
\end{equation}

Note that the in-plane displacements ($u^{(m)}$ and $v^{(m)}$) are expressed in terms of the transverse deflection ($w^{(m)}$) of the reference surface and exhibit linear variations with respect to the $z$ coordinate. Subsequently, the strain components can be expressed in terms of the displacement components as, 
\begin{equation}
    \mathbf{S}^{(m)}= \{\varepsilon_{x}^{(m)}, \varepsilon_{y}^{(m)}, 2\varepsilon_{xy}^{(m)}\}^{T} = -z \left\{\dfrac{\partial^2 w^{(m)}}{\partial x^2}, \dfrac{\partial^2 w^{(m)}}{\partial y^2}, 2\dfrac{\partial^2 w^{(m)}}{\partial x \partial y} \right\}^{T}
\end{equation}

The constitutive relation of an isotropic material under plane-stress conditions relates the stress components $\mathbf{T}^{(m)} =  \{\sigma_{x}^{(m)}, \sigma_{y}^{(m)}, \sigma_{xy}^{(m)}\}^{T}$ to the strain components $\mathbf{S}^{(m)} =  \{\varepsilon_{x}^{(m)}, \varepsilon_{y}^{(m)}, 2\varepsilon_{xy}^{(m)}\}^{T}$ in a two-dimensional system can be expressed mathematically as, 
\begin{equation}
    \label{T}
    \mathbf{T}^{(m)} = \mathbf{C_s} \mathbf{S}^{(m)},
\end{equation}
where $\mathbf{C_s}$ is the elastic stiffness matrix, which characterises the linear relationship between the stress and strain components. It is expressed in terms of Young's modulus ($E_s$) and Poisson's ratio ($\nu_s$) using the following equation:
\begin{equation}
    \mathbf{C_s} = \frac{E_s}{1 - \nu_s^2} 
                   \begin{bmatrix} 
                   1 & \nu_s & 0 \\
                   \nu_s & 1 & 0 \\
                   0 & 0 & \frac{1 - \nu_s}{2} \\
                   \end{bmatrix}.
\end{equation}

Meanwhile, for piezoelectric layers polarised in the $z$-direction, the constitutive equations that establish the relationship between the stress components $\mathbf{T}^{(m)} = \{\sigma_{x}^{(m)}, \sigma_{y}^{(m)}, \sigma_{xy}^{(m)}\}^{T}$ and the electric displacement $\mathbf{D}^{(m)} = \{0, 0, D_z^{(m)}\}$ in relation to the strain components $\mathbf{S}^{(m)} = \{\varepsilon_{x}^{(m)}, \varepsilon_{y}^{(m)}, 2\varepsilon_{xy}^{(m)}\}^{T}$ and the electric field $\mathbf{E}^{(m)} = \{0, 0, E_z^{(m)}\}^{T}$ can be expressed as,
\begin{equation}
    \label{T-D}
   \begin{Bmatrix}
   \mathbf{T}^{(m)} \\
   \mathbf{D}^{(m)}
   \end{Bmatrix} =
                   \begin{bmatrix} 
                   \mathbf{C}_p^E & -\mathbf{e}^{T} \\
                   \mathbf{e} & \boldsymbol{\varepsilon^S} \\
                   \end{bmatrix}
   \begin{Bmatrix}
   \mathbf{S}^{(m)} \\
   \mathbf{E}^{(m)}
   \end{Bmatrix}
\end{equation}
where, the matrices $\mathbf{C_p}$, $\mathbf{e}$, and $\boldsymbol{\varepsilon^S}$ define the linear relationship characterising transversely isotropic material behaviour within the $xy$-plane. The plain-stress condition in the piezoelectric layer results in,
\begin{equation}
    \label{cpE-e-epsilon}
    \mathbf{C}_p^E =   \begin{bmatrix} 
                          \bar{c}_{11}^E & \bar{c}_{11}^E & 0 \\
                          \bar{c}_{12}^E & \bar{c}_{22}^E & 0 \\
                          0 & 0 & \bar{c}_{66}^E\\
                         \end{bmatrix}, \,\,\,
    \mathbf{e} =   \{\bar{e}_{31}, \bar{e}_{32}, 0\}^T, \,\,\, \boldsymbol{\varepsilon}^S = \bar{\varepsilon}_{33}{^S}
\end{equation}

Based on Hamilton's principle, the weak formulation of the problem can be expressed as follows \cite{crandall1968dynamics},
\begin{equation}
    \label{weakform}
\begin{split} 
    \int_{t_1}^{t_2}&\left[ \delta(T^{(m)}-V^{(m)}+W_e^{(m)}) +\delta W^{(m)} \right] dt= 0
\end{split}
\end{equation}
where the variation of the kinetic energy ($\delta T^{(m)}$), the variation of the elastic potential energy ($\delta V^{(m)}$), the variation of the electrical potential energy ($\delta W_e^{(m)}$) and the variation of the non-conservative work ($\delta W^{(m)}$) are given by, 
\begin{equation}
\begin{split} 
    \delta T^{(m)} &= \int_{V_s^{(m)}}\rho_s\delta\dot{\textbf{u}}^{(m)T}\dot{\textbf{u}}^{(m)}dV_s^{(m)}  + \int_{V_p^{(m)}}\rho_p\delta\dot{\textbf{u}}^{(m)T}\dot{\textbf{u}}^{(m)}dV_p^{(m)} \\
    &+ \sum_{i=1}^{n_m^{(m)}} \delta\dot{\textbf{u}}^{(m)T}(x_i,y_i,t) \ \dot{\textbf{u}}^{(m)}(x_i,y_i,t)\ m_i^{(m)}(x_j,y_j,t)\\
    \delta V^{(m)} &= \int_{V_s^{(m)}}\delta\textbf{S}^{(m)T}\textbf{T}^{(m)}dV_s^{(m)}  + \int_{V_p^{(m)}}\delta\textbf{S}^{(m)T}\textbf{T}^{(m)}dV_p^{(m)} \\
    \delta W_e^{(m)} &= \int_{V_p^{(m)}}\delta\textbf{E}^{(m)T}\textbf{D}^{(m)}dV_p^{(m)} \\
    \delta W^{(m)} &= \sum_{j=1}^{n_f^{(m)}} \delta\textbf{u}^{(m)}(x_j,y_j,t) \textbf{f}_j^{(m)}+ \sum_{k=1}^{n_q^{(m)}} \delta\varphi^{(m)}(x_k,y_k,t) q^{(m)}(x_k,y_k,t)
\end{split}
\end{equation}
In the variational kinetic energy expression, it is considered a set of point masses $\{m_i^{(m)},\ i=1...n_m^{(m)}\}$ situated at coordinates ($x_i$, $y_i$). On the other hand, in the variational formulation of the non-conservative expression, it is incorporated a set of discrete mechanical forces $\{\textbf{f}_j^{(m)},\ j=1...n_f^{(m)}\}$ applied at coordinates ($x_j$, $y_j$), and discrete electric charge outputs $q^{(m)}$ extracted into the system. Here, $\varphi^{(m)}$ represents the scalar electrical potential.

In the specific case of elastic potential energy, it can be integrated within the domain $\Omega^{(m)}$ and on the boundary $\Gamma^{(m)}$, and subsequently expressed as follows \cite{bittencourt2014computational}:
\begin{equation}
    \begin{split} 
    V^{(m)} &= \int_{\Omega^{(m)}} q_i^{(m)} w^{(m)} d\Omega^{(m)} + \int_{\Gamma^{(m)}} (Q_x^{(m)}n_x^{(m)}+Q_y^{(m)}n_y^{(m)})w^{(m)} d\Gamma^{(m)} \\
    &- \int_{\Gamma^{(m)}} (M_{xx}^{(m)}n_x^{(m)}+M_{xy}^{(m)}n_y^{(m)})\dfrac{\partial w^{(m)} }{\partial x} d\Gamma^{(m)} -  \int_{\Gamma^{(m)}} (M_{xy}^{(m)}n_x^{(m)}+M_{yy}^{(m)}n_y^{(m)})\dfrac{\partial w^{(m)} }{\partial y} d\Gamma^{(m)} 
    \end{split}
\end{equation}
where the vector of the moment is given by, 
\begin{equation}
    \{M_{xx}^{(m)}, M_{yy}^{(m)}, M_{xy}^{(m)}\}^{T} = \int_{-t_s/2}^{t_s/2} z \mathbf{C_s} \mathbf{S}^{(m)} dz + 2\int_{-t_s/2-t_p}^{t_s/2+t_p} z \mathbf{C_p} \mathbf{S}^{(m)} dz - 2\int_{-t_s/2-t_p}^{t_s/2+t_p} z  \mathbf{e} \mathbf{E}^{(m)} dz ,
\end{equation}
the shear force can be expressed as, 
\begin{equation}
    \label{D-E}
    Q_x^{(m)} = \dfrac{\partial M_{xx}^{(m)} }{\partial x} + \dfrac{\partial M_{xy}^{(m)} }{\partial y} , \,\,\, Q_y^{(m)} = \dfrac{\partial M_{yy}^{(m)} }{\partial y} + \dfrac{\partial M_{xy}^{(m)} }{\partial x}
\end{equation}
and $q_i^{m}$ corresponds to the internally distributed load, formulated as follows: 
\begin{equation}
    q_i^{(m)} = \dfrac{\partial^2 M_{xx}^{(m)} }{\partial x^2} + 2\dfrac{\partial^2 M_{xy}^{(m)} }{\partial x \partial y} + \dfrac{\partial^2 M_{yy}^{(m)} }{\partial y^2}
\end{equation}

The Nitsche extension's consistency terms for a Kirchhoff-Love patch coupling on the boundary $\Gamma^*$ given by \cite{guo2016isogeometric}

\footnotesize
\begin{equation}
    \label{nitschet_contri}
    \begin{split} 
    V^{NIT}_{CS} = \int_{\Gamma^*} \delta \langle Q_xn_x+Q_yn_y \rangle [\![ w ]\!] d\Gamma^* - \int_{\Gamma^*} \delta \langle M_{xx}n_x+M_{xy}n_y \rangle \left[\!\left[ \dfrac{\partial w }{\partial x} \right]\!\right] d\Gamma^* -  \int_{\Gamma^*} \delta \langle M_{xy}n_x+M_{yy}n_y \rangle\left[\!\left[ \dfrac{\partial w }{\partial y} \right]\!\right] d\Gamma^*
    \end{split}
\end{equation}
\normalsize

where the double square brackets denote jump operator
\begin{equation}
    [\![ w ]\!] = w^{(1)}-w^{(2)}, \,\,\, \left[\!\left[ \dfrac{\partial w }{\partial x} \right]\!\right] = \dfrac{\partial w^{(1)} }{\partial x} - \dfrac{\partial w^{(2)} }{\partial x}, \,\,\, \left[\!\left[ \dfrac{\partial w }{\partial y} \right]\!\right] = \dfrac{\partial w^{(1)} }{\partial y} - \dfrac{\partial w^{(2)} }{\partial y}
\end{equation}
and the angle bracket denote the average operators
\begin{equation}
\label{angleBracket}
\begin{split} 
    \langle Q_xn_x+Q_yn_y \rangle &= \gamma( Q_x^{(1)}n_x+Q_y^{(1)}n_y ) + (1-\gamma)( Q_x^{(2)}n_x+Q_y^{(2)}n_y ) \\
    \langle M_{xx}n_x+M_{xy}n_y \rangle &= \gamma( M_{xx}^{(1)}n_x+M_{xy}^{(1)}n_y ) + (1-\gamma)( M_{xx}^{(2)}n_x+M_{xy}^{(2)}n_y )  \\
    \langle M_{xy}n_x+M_{yy}n_y \rangle &= \gamma( M_{xy}^{(1)}n_x+M_{yy}^{(1)}n_y ) + (1-\gamma)( M_{xy}^{(2)}n_x+M_{yy}^{(2)}n_y ) \\
\end{split}
\end{equation}\\

The value of weighting parameter $\gamma$ in Equation \ref{angleBracket} determines the relative contribution of the two coupled domains, $\Omega^1$ and $\Omega^2$, to enforce the compatibility condition.

Stability terms of the weak coupling extension are included. These terms are expressed in terms of the neutral plane displacement $w$ and its derivatives \cite{guo2016isogeometric},

\begin{equation}
\label{Nit_St}
    V^{NIT}_{ST} = \alpha \int_{\Gamma^*} t \delta [\![ w ]\!] \cdot [\![ w ]\!] d\Gamma^* + \alpha \int_{\Gamma^*} \frac{t^3}{12} \delta \left[\!\left[ \dfrac{\partial w }{\partial x} \right]\!\right] \cdot \left[\!\left[ \dfrac{\partial w }{\partial x} \right]\!\right] d\Gamma^* + \alpha \int_{\Gamma^*} \frac{t^3}{12} \delta \left[\!\left[ \dfrac{\partial w }{\partial y} \right]\!\right] \cdot \left[\!\left[ \dfrac{\partial w }{\partial y} \right]\!\right] d\Gamma^*
\end{equation}

The stabilisation parameter $\alpha$, which is included as a penalty term, serves to enforce the coercivity of the Nitsche-type bilinear form.

Hence, the weak form for the Piezoelectric Energy Harvester (PEH) is explicitly formulated, taking into account Nitsche-based conditions for the patch interface. It’s important to note that while certain terms were initially introduced for the scenario where the board is modelled with only two patches, these terms can readily be extended to accommodate cases involving three or more patches.

\section{Isogeometric analysis (IGA)}
\label{S:3}
In Isogeometric Analysis (IGA), the description of the geometry patch $\mathbf{x}^{(m)} = (x,y)\in\Omega^{(m)}$  and the approximation of the unknown displacement $w^{(m)}$ are achieved using two-dimensional NURBS basis functions $N_I^{(m)}$. NURBS are explained in Appendix A. This can be expressed as:
\begin{equation}
    \label{Eq.x}
    \textbf{x}^{(m)}(\boldsymbol{\xi}^{(m)})=\sum^{k^{(m)}}_{I=1}N_I^{(m)}(\boldsymbol{\xi}^{(m)})\tilde{\textbf{x}}^{(m)}_I,
\end{equation}
\begin{equation}
    \label{Eq.ww}
    w^{(m)}(\boldsymbol{\xi}^{(m)}, t)=\sum_{I=1}^{k^{(m)}}N_I^{(m)}(\boldsymbol{\xi}^{(m)})w_I^{(m)}(t),
\end{equation}
where, each element is characterised by a specific number of basis functions, symbolised by $k^{(m)}$. The so-called control variables, $w_I^{(m)}(t)$, represent the deflection projected at control point I. Meanwhile, the coordinates $\boldsymbol{\xi}^{(m)}$ denote the parameter space. The study investigates two scenarios for connecting patches. In the first scenario, all patches are connected via a common electrode, forming a single electrical circuit. Alternatively, in the second scenario, each patch has its own independent electrode, connecting it to a separate circuit. We will explore the shared electrode configuration before discussing the case of independent electrodes. 

\subsection{Single electrical circuit}
In this configuration, the piezoelectric device is connected to a single electrical resistance ($R_l$), with the generated voltage denoted as $v_p$. By substituting Equations \ref{Eq.x} and \ref{Eq.ww} into the variational formulation (as defined in Equation \ref{weakform}), it is obtained a discrete system of equations in a straightforward manner. These discrete equations can be expressed in a Nitsche-type form, considering the displacement vector $\textbf{w} = \{\textbf{w}^{(1)}\quad \textbf{w}^{(2)} \}^T$, which contains the displacement in the $K$ control points in both domains in the two-patch scenario described in Section \ref{S:3}.

\begin{equation}
\label{eq:MECH}
      \textbf{M}\ddot{\textbf{w}}+\textbf{C}\dot{\textbf{w}}+
    \textbf{K}\textbf{w} - \boldsymbol{\Theta}v_p  =-\textbf{M}\textbf{r}\,a_b\\
\end{equation}
\begin{equation}
    C_p \dot{v_p} + \frac{v_p}{R_l} + \boldsymbol{\Theta}^{T}\textbf{w} =0 
\label{eq:ELEC}
\end{equation}
The definition of the system's global mass matrix $\textbf{M}\in\mathbb{R}^{K \times K}$, the system's global stiffness matrix $\textbf{K}\in\mathbb{R}^{K \times K}$, the system's global electromechanical coupling vector $\boldsymbol{\Theta}\in\mathbb{R}^{K \times 1}$ and the system's global capacitance $C_p\in\mathbb{R}$ are provided below,
\begin{equation}
\label{M_Eq}
\textbf{M} = 
\begin{bmatrix}
\textbf{M}^{(1)}    & \mathbf{0} \\
\mathbf{0}        & \textbf{M}^{(2)} 
\end{bmatrix}
\end{equation}

\begin{equation}
\label{K_global_Eq}
\textbf{K} = 
\begin{bmatrix}
\textbf{K}^{(1)}+\textbf{K}^{(1)}_n+\textbf{K}^{(1)}_s    & \textbf{K}_c \\
\textbf{K}^T_c        & \textbf{K}^{(2)}+\textbf{K}_n^{(2)}+\textbf{K}_s^{(2)} 
\end{bmatrix}
\end{equation}

\begin{equation}
\label{theta_global_Eq}
\boldsymbol{\Theta} = 
\begin{Bmatrix}
\boldsymbol{\Theta}^{(1)} + \boldsymbol{\Theta}^{(1)}_n\\
\boldsymbol{\Theta}^{(2)} + \boldsymbol{\Theta}^{(2)}_n
\end{Bmatrix}
\end{equation}

\begin{equation}
\label{cP_Eq}
C_p = \sum_{m=1}^2 C_p^{(m)}
\end{equation}
Other significant parameters include the vector of ones $\textbf{r} = \{1,...,1\} \in\mathbb{R}^{K \times 1}$, and the global mechanical damping matrix $\textbf{C} = \alpha_o \textbf{M} + \beta_o \textbf{K} \in\mathbb{R}^{K \times K}$, which is assumed to be proportional to the mass and stiffness matrices.

It's important to note that the system's global matrices are assembled from the global matrices of each individual patch. These include the global mass matrix $\textbf{M}^{(m)}\in\mathbb{R}^{K^{(m)} \times K^{(m)}}$, the global stiffness matrix $\textbf{K}^{(m)}\in\mathbb{R}^{K^{(m)} \times K^{(m)}}$, the global electromechanical coupling matrix $\boldsymbol{\Theta}^{(m)}\in\mathbb{R}^{K \times 1}$, and the capacitance $C_p^{(m)}\in\mathbb{R}$. In this context, $K^{(m)}$ denotes the number of control points in each patch. The assembly of these global matrices is achieved through the integration of the following sub-matrices:

\begin{equation}
\label{Mi_Eq}
      M^{(m)}_{IJ} = \rho_s\int_{\Omega^{(m)}} \textbf{R}_I^{(m)T}\textbf{m}_s\textbf{R}^{(m)}_J d\Omega^{(m)} + 2\rho_p\int_{\Omega^{(m)}} \textbf{R}_I^{(m)T}\textbf{m}_p\textbf{R}^{(m)}_J d\Omega^{(m)}
\end{equation}

\begin{equation}
\label{Ki_Eq}
      K^{(m)}_{IJ} = \frac{t_s^3}{12}\int_{\Omega^{(m)}} \textbf{B}_I^{(m)T}\textbf{C}_s\textbf{B}^{(m)}_J d\Omega^{(m)} + \left( \frac{2t_p^3}{3}+t_p^2t_s+\frac{t_s^2}{2} \right) \int_{\Omega^{(m)}} \textbf{B}_I^{(m)T}\textbf{C}_p\textbf{B}^{(m)}_J d\Omega^{(m)}
\end{equation}
\begin{equation}
\label{Theta_Eq}
      \Theta^{(m)}_{I} = (t_p+t_s)\int_{\Omega^{(m)}} \textbf{e}^T\textbf{B}^{(m)}_I d\Omega^{(m)} 
\end{equation}

\begin{equation}
    C_p^{(m)} = 2\int_{\Omega^{(m)}}\frac{\bar{\varepsilon}_{33}^S }{t_p} d\Omega^{(m)}
    \label{eq:Cp}
\end{equation}
where,

\begin{equation}
\label{BI}
      \textbf{B}_I^{(m)} = \{ -N_{I,xx}^{(m)} \quad -N_{I,yy}^{(m)} \quad -2N_{I,xy}^{(m)} \}^T
\end{equation}

\begin{equation*}
\label{ms_Eq}
\textbf{m}_s = 
\begin{bmatrix}
t_s  & 0        &  0\\
0    & t_s^3/12 &  0\\
0    & 0        &  t_s^3/12
\end{bmatrix}
; \quad 
\textbf{m}_p = 
\begin{bmatrix}
t_p  & 0                                    &  0\\
0    & t_p^3/3 + t_p^2t_s/2 + t_pt_s^2/4    &  0\\
0    & 0                                    &  t_p^3/3 + t_p^2t_s/2 + t_pt_s^2/4
\end{bmatrix}
\end{equation*}

\begin{equation}
\label{RI}
      \textbf{R}_I^{(m)} = \{ N_{I}^{(m)} \quad N_{I,x}^{(m)} \quad N_{I,y}^{(m)} \}^T
\end{equation}

A thorough comprehension of the system's global stiffness matrix in Equation \ref{K_global_Eq} requires a detailed explanation of its constituent parts. This matrix comprises several matrices, including the bulk global stiffness matrices $\textbf{K}^{(m)}$, defined in Equation \ref{Ki_Eq}. $\textbf{K}^{(m)}_n$ denote the Nitsche contribution matrix, which is derived from the first two terms in Equation \ref{nitschet_contri}. The matrix is assembled from the following sub-matrices assuming the weighting parameter $\gamma$ = 0.5,

\begin{equation}
\label{Kn_Eq}
\begin{split}
      K^{(m)}_{nIJ} = \frac{1}{2} \frac{t_s^3}{12} &\left( \int_{\Gamma^*} \textbf{R}_I^{(m)T}\textbf{n}_1\textbf{C}_s\textbf{B}^{(m)}_J d\Gamma^* + \int_{\Gamma^*} \textbf{B}^{(m)T}_J\textbf{C}_s^T\textbf{n}_1^T\textbf{R}_I^{(m)} d\Gamma^* \right) + ... \\ \frac{1}{2}\left( \frac{2t_p^3}{3}+t_p^2t_s+\frac{t_s^2}{2} \right) & \left( \int_{\Gamma^*} \textbf{R}_I^{(m)T}\textbf{n}_1\textbf{C}_p\textbf{B}^{(m)}_J d\Gamma^* + \int_{\Gamma^*} \textbf{B}^{(m)T}_J\textbf{C}_s^T\textbf{n}_1^T\textbf{R}_I^{(m)} d\Gamma^* \right) + ... \\
      \frac{1}{2} \frac{t_s^3}{12}& \left( \int_{\Gamma^*} \textbf{R}_I^{(m)T}\textbf{n}_2\textbf{C}_s\textbf{B}^{(m)}_{J,x} d\Gamma^* + \int_{\Gamma^*} \textbf{R}_I^{(m)T}\textbf{n}_3\textbf{C}_s\textbf{B}^{(m)}_{J,y} d\Gamma^* \right)  + ...\\
      \frac{1}{2} \frac{t_s^3}{12}& \left( \int_{\Gamma^*} \textbf{B}^{(m)T}_{J,x}\textbf{C}_s^T\textbf{n}_2^T\textbf{R}_I^{(m)} d\Gamma^* + \int_{\Gamma^*} \textbf{B}^{(m)T}_{J,y}\textbf{C}_s^T\textbf{n}_3^T\textbf{R}_I^{(m)} d\Gamma^* \right) + ...\\
      \frac{1}{2} \left( \frac{2t_p^3}{3}+t_p^2t_s+\frac{t_s^2}{2} \right)& \left( \int_{\Gamma^*} \textbf{R}_I^{(m)T}\textbf{n}_2\textbf{C}_p\textbf{B}^{(m)T}_{J,x} d\Gamma^* + \int_{\Gamma^*} \textbf{R}_I^{(m)T}\textbf{n}_3\textbf{C}_p\textbf{B}^{(m)}_{J,y} d\Gamma^* \right)  + ...\\
      \frac{1}{2} \left( \frac{2t_p^3}{3}+t_p^2t_s+\frac{t_s^2}{2} \right)& \left( \int_{\Gamma^*} \textbf{B}^{(m)T}_{J,x}\textbf{C}_p^T\textbf{n}_2^T\textbf{R}_I^{(m)} d\Gamma^* + \int_{\Gamma^*} \textbf{B}^{(m)T}_{J,y}\textbf{C}_p^T\textbf{n}_3^T\textbf{R}_I^{(m)} d\Gamma^* \right) \\
\end{split}
\end{equation}
where,
\begin{equation}
    \textbf{n}_1 = 
    \begin{bmatrix}
    0    & 0        &  0\\
    n_x  & 0        &  n_y\\
    0    & n_y        &  n_x
    \end{bmatrix}
    \quad \textbf{n}_2 =
    \begin{bmatrix}
    n_x  & 0        &  n_y\\
    0    & 0        &  0\\
    0    & 0        &  0
    \end{bmatrix}
    \quad \textbf{n}_3 =
    \begin{bmatrix}
    n_y  & 0        &  n_x\\
    0    & 0        &  0\\
    0    & 0        &  0
    \end{bmatrix}
\end{equation}

While the stabilization matrix, denoted as $\textbf{K}^i_s$, is derived from Equation \ref{Nit_St} and built from the assembling of the following submatrix,

\begin{equation}
\label{Ks_Eq}
      K^{(m)}_{sIJ} = \alpha \left( \int_{\Gamma^*} \textbf{R}_I^{(m)T}\textbf{m}_s\textbf{R}^{(m)}_J d\Gamma^* + 2\int_{\Gamma^*} \textbf{R}_I^{(m)T}\textbf{m}_p\textbf{R}^{(m)}_J d\Gamma^* \right)
\end{equation}

Furthermore, the symmetrical matrix $\textbf{K}_c$ represents the coupling matrix and is derived from Equations \ref{Nit_St} and \ref{Ki_Eq}. It is assembled as follows,

\begin{equation}
\label{Kc_Eq}
\begin{split}
      K_{cIJ} = \alpha &\left( \int_{\Gamma^*} \textbf{R}_I^{(1)T}\textbf{m}_s\textbf{R}^{(2)}_J d\Gamma^* + 2\int_{\Gamma^*} \textbf{R}_I^{(1)T}\textbf{m}_p\textbf{R}^{(2)}_J d\Gamma^* \right)  +...\\ \frac{1}{2} \frac{t_s^3}{12} & \left( \int_{\Gamma^*} \textbf{R}_I^{(1)T}\textbf{n}_1\textbf{C}_s\textbf{B}^{(2)}_J d\Gamma^* + \int_{\Gamma^*} \textbf{B}^{(1)T}_J\textbf{C}_s^T\textbf{n}_1^T\textbf{R}_I^{(2)} d\Gamma^* \right)+ ... \\ \frac{1}{2} \left( \frac{2t_p^3}{3}+t_p^2t_s+\frac{t_s^2}{2} \right) & \left( \int_{\Gamma^*} \textbf{R}_I^{(1)T}\textbf{n}_1\textbf{C}_p\textbf{B}^{(2)}_J d\Gamma^* + \int_{\Gamma^*} \textbf{B}^{(1)T}_J\textbf{C}_p^T\textbf{n}_1^T\textbf{R}_I^{(2)} d\Gamma^* \right)+ ...\\
      \frac{1}{2} \frac{t_s^3}{12}& \left( \int_{\Gamma^*} \textbf{R}_I^{(1)T}\textbf{n}_2\textbf{C}_s\textbf{B}^{(2)}_{J,x} d\Gamma^* + \int_{\Gamma^*} \textbf{R}_I^{(1)T}\textbf{n}_3\textbf{C}_s\textbf{B}^{(2)}_{J,y} d\Gamma^* \right)  + ...\\
      \frac{1}{2} \frac{t_s^3}{12}& \left( \int_{\Gamma^*} \textbf{B}^{(1)T}_{J,x}\textbf{C}_s^T\textbf{n}_2^T\textbf{R}_I^{(2)} d\Gamma^* + \int_{\Gamma^*} \textbf{B}^{(1)T}_{J,y}\textbf{C}_s^T\textbf{n}_3^T\textbf{R}_I^{(2)} d\Gamma^* \right) + ...\\
      \frac{1}{2} \left( \frac{2t_p^3}{3}+t_p^2t_s+\frac{t_s^2}{2} \right)& \left( \int_{\Gamma^*} \textbf{R}_I^{(1)T}\textbf{n}_2\textbf{C}_p\textbf{B}^{(2)}_{J,x} d\Gamma^* + \int_{\Gamma^*} \textbf{R}_I^{(1)T}\textbf{n}_3\textbf{C}_p\textbf{B}^{(2)}_{J,y} d\Gamma^* \right)  + ...\\
      \frac{1}{2} \left( \frac{2t_p^3}{3}+t_p^2t_s+\frac{t_s^2}{2} \right)& \left( \int_{\Gamma^*} \textbf{B}^{(1)T}_{J,x}\textbf{C}_p^T\textbf{n}_2^T\textbf{R}_I^{(2)} d\Gamma^* + \int_{\Gamma^*} \textbf{B}^{(1)T}_{J,y}\textbf{C}_p^T\textbf{n}_3^T\textbf{R}_I^{(2)} d\Gamma^* \right) \\
\end{split}
\end{equation}

Similarly, to comprehend the global electromechanical coupling in Equation \ref{theta_global_Eq}, the Nitsche contribution vector $\Theta_n^{(m)}$ is introduced. This vector is derived from the second and third terms in Equation \ref{nitschet_contri} and is constructed by assembling the following matrix:

\begin{equation}
\label{Theta_nit_Eq}
      \Theta^{(m)}_{nI} = (t_p+t_s)\int_{\Omega^{(m)}} \textbf{n}_4\textbf{e}^T\textbf{B}^{(m)}_I d\Omega^{(m)} 
\end{equation}
where,
\begin{equation}
    \textbf{n}_4 = 
    \begin{bmatrix}
    n_x  & 0        &  0\\
    0    & n_y      &  0\\
    0    & 0        &  0
    \end{bmatrix}
\end{equation}

Therefore, by considering that the base acceleration in the transverse direction is a harmonic motion expressed as $a_b = A_Be^{j \omega t}$, it is possible to define the output voltage as $v_p = V_pe^{j\omega t}$, where $j = \sqrt{-1}$. Using Equations \ref{eq:MECH} and \ref{eq:ELEC}, the Frequency Response Function (FRF) that links the output voltage $V_p = V_p(\omega)$ and the base acceleration $A_p = A_p(\omega)$ for given excitation frequency $\omega$ can be derived as follow: 

\small
\begin{equation}
    H(\omega) = \frac{V_P( \omega)}{A_B( \omega)}=j\omega\left(\frac{1}{R_l}+j\omega C_p\right)^{-1} \boldsymbol{\Theta}^T\left(-\omega^2\textbf{M}+j\omega \textbf{C} + \textbf{K} + j\omega \left(\frac{1}{R_l}+j\omega C_p\right)^{-1} \boldsymbol{\Theta} \boldsymbol{\Theta}^T\right)^{-1}\textbf{M}\textbf{r}
\end{equation}
\normalsize

\subsection{Independent electrical circuits}
As previously discussed, the study explores two scenarios for connecting patches. In this sense, we delve into the concept of an independent electrode configuration. Specifically, each patch is connected to an individual electrical resistance denoted as $R_l^{(m)}$. The voltage generated by each patch is represented as $v_p^{(m)}$. Consequently, the discrete equations describing the system can be expressed as follows:

\begin{equation}
\label{eq:mech_02}
\textbf{M}\ddot{\textbf{w}}+\textbf{C}\dot{\textbf{w}}+ \textbf{K}\textbf{w} - \sum_{m=1}^2 \boldsymbol{\Theta}_G^{(m)}v_p^{(m)}  =-\textbf{M}\textbf{r}\,a_b 
\end{equation}
\begin{equation}    
\label{eq:elec_02}
C_p^{(m)} \dot{v_p}^{(m)} + \frac{ v_p^{(m)} }{ R_l^{(m)} } +\boldsymbol{\Theta}^{(m)T}_G\textbf{w} = 0 
\end{equation}
Here, the equation that characterises the device’s mechanical behaviour is derived, incorporating electrical coupling. Furthermore, it presents the equation for the electrical circuit considering the mechanical coupling pertinent to each patch. It is practical to redefine the system’s electromechanical coupling vector, denoted as $\boldsymbol{\Theta}$, into separate $\boldsymbol{\Theta}_G^{(m)}$ vectors for individual patches. The equation below presents the vector for the case of two patches:

\begin{equation}
\label{theta_global_Eq_2}
\boldsymbol{\Theta}_G^{(1)} = 
\begin{Bmatrix}
\boldsymbol{\Theta}^{(1)}\\
\boldsymbol{0}
\end{Bmatrix}
\quad
\textrm{and}
\quad
\boldsymbol{\Theta}_G^{(2)} = 
\begin{Bmatrix}
\mathbf{0}\\
\boldsymbol{\Theta}^{(2)} 
\end{Bmatrix}
\end{equation}

Following on the premise that the base transverse acceleration is harmonic, we derive the FRF. This function correlates the output voltage amplitude for each patch with the amplitude of the base acceleration. The derivation is based on Equations \ref{eq:mech_02} and \ref{eq:elec_02}, leading to an expression for the FRF as follows
\footnotesize
\begin{equation}
    H^{(m)}(\omega) =j\omega\left(\frac{1}{R_l^{(m)}}+j\omega C_p^{(m)}\right)^{-1} \boldsymbol{\Theta}_G^{(m)T}\left(-\omega^2\textbf{M}+j\omega \textbf{C} + \textbf{K} + \sum_{m=1}^2 j\omega \left(\frac{1}{R_l^{(m)}}+j\omega C_p^{(m)}\right)^{-1} \boldsymbol{\Theta}_G^{(m)} \boldsymbol{\Theta}_G^{(m)T}\right)^{-1}\textbf{M}\textbf{r}
\end{equation}
\normalsize

\section{Validation of the multi-patch IGA model}
\label{S:4}
In this section, we present a series of numerical examples of Kirchhoff-Love multi-patch plates to demonstrate the effectiveness and accuracy of the PEH approach in addressing a diverse range of geometric configurations. The results of these examples are also compared with those found in the existing literature to provide further validation.

\subsection{Simply supported plate under sinusoidal distributed load}
In the first example, a plate with a sinusoidal distributed load over its surface is considered to study the accuracy of the multi-patch model. The plate is a square with a length $a$=$b$=10 m and thickness $t$=0.01m while the material properties are Young’s modulus $E$ = 100MPa, Poisson’s ratio $\nu$ = 0.3. The boundary conditions on the four edges are simple supports and a sinusoidal distributed load is defined as, 
\begin{equation}
q(x,y) = q_0 \sin \frac{\pi x}{a} \sin \frac{\pi y}{b}
\end{equation}
where $q_0$=-10N. The Navier solution for the exact deflection is expressed as \cite{reddy2006theory}, 
\begin{equation}
\label{wexact}
w^{\text{exact}} = \frac{q_0}{\pi^4 D_0 \left(\frac{1}{a^2}+\frac{1}{b^2}\right)^2} \sin \frac{\pi x}{a} \sin \frac{\pi y}{b}
\end{equation}

\begin{figure}[ht]
	\centering
	\includegraphics[width=0.3333\textwidth]{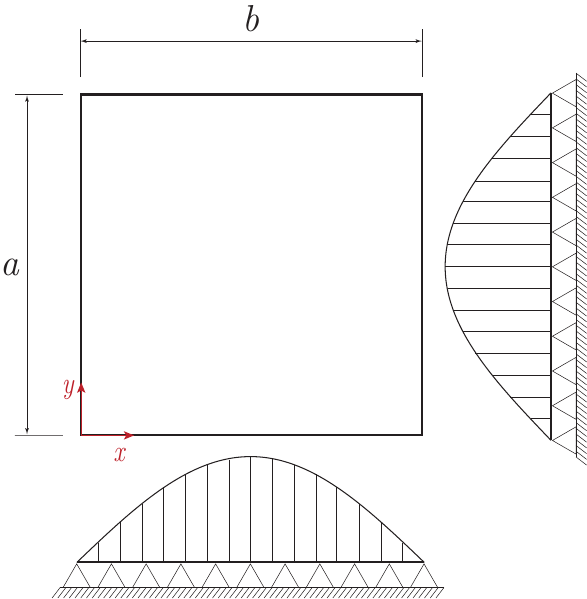}
\caption{Scheme of a simply supported square plate under a sinusoidal load.}
	\label{fig:caseA}
\end{figure}

\begin{figure}[ht]
	\centering
	\includegraphics[width=1\textwidth]{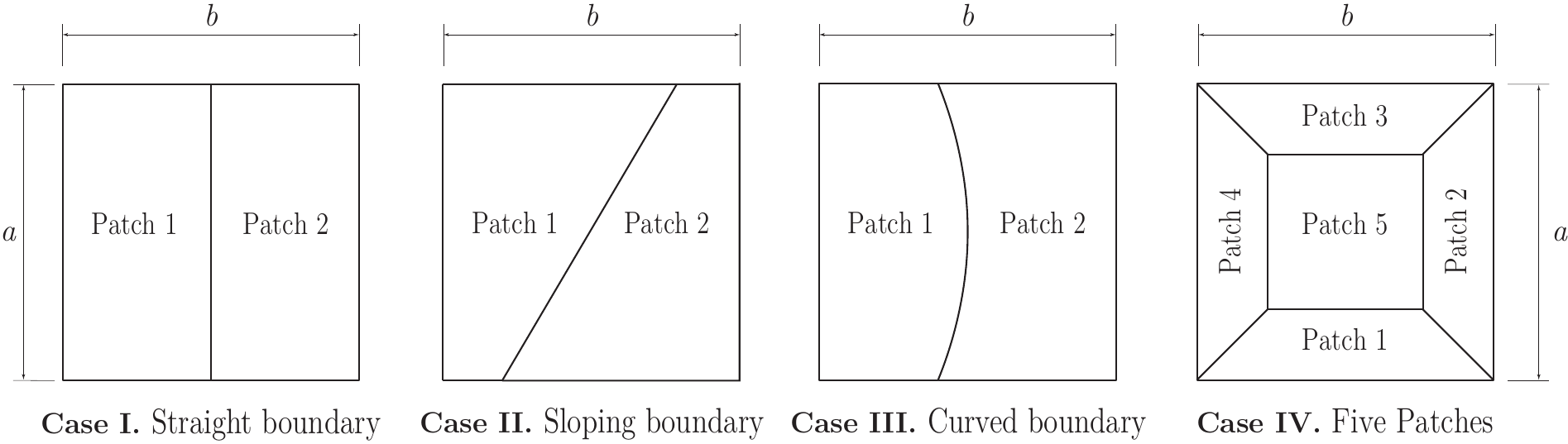}
\caption{Cases studied for the construction of the simply supported square plate and the rectangular piezoelectric harvester.}
	\label{Ch6:fig:caseA}
\end{figure}

Note that in this example, we consider only a Kirchhoff-Love plate without piezoelectric layers. Four different approaches for constructing the square plate using multi-patches were analysed to evaluate the validity of the Nitsche method on various interface shapes between patches. These cases are illustrated in Figure \ref{Ch6:fig:caseA}. The \textbf{Case I} comprises two identical NURBS patches, where the boundary is aligned with the $y$-axis. The \textbf{Case II} also employs two patches, but the interface has an inclination of 30 degree with respect to the $y$-axis, passing through the centre of the plate. The \textbf{Case III} involves a boundary parametrised by quadratic B-Splines, with control points located at ($0.4b$, $0$), ($0.6 b$, $0.5 a$) and ($0.4 b$, $a$). The \textbf{Case IV} employs five NURBS patches to assemble the structure. One of them is a square patch with a side length of $a/2=b/2$, located in the centre of the plate. The other four patches have a trapezoidal shape surrounding the square patch, with dimensions of a high $a/2=b/2$ and bases $a=b$ and $a/2=b/2$. Furthermore, a reference case is considered in which the plate is constructed using a single patch.

The accuracy of the numerical solution $w^h$ in the four cases described above is compared to the solution obtained by modelling the plate with a single patch. The accuracy is estimated with respect to the analytical solution $w^{\text{exact}}$ of Equation \ref{wexact} using the relative error $L2$-norm defined as follows,
\begin{equation}
     e = \frac{\lVert w^{\text{exact}} - w^h \rVert_{L_2(\Omega)}}{\lVert w^{\text{exact}} \rVert_{L_2(\Omega)}}
\end{equation}

\begin{figure}[ht]
	\centering
	\includegraphics[width=1.0\textwidth]{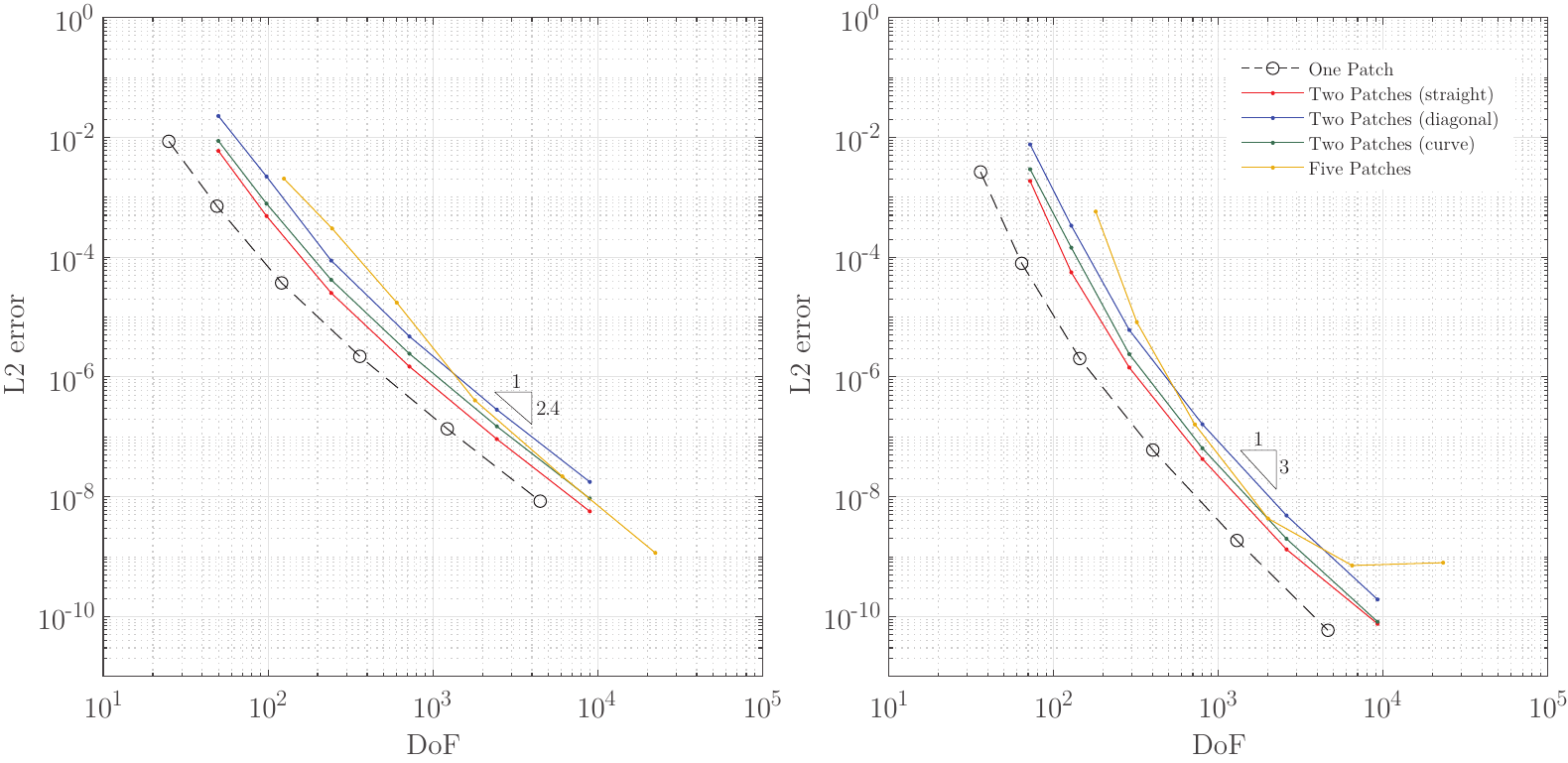}
\caption{Comparison of multi-patches IGA solutions in terms of $L_2$ error vs. the number of degrees of freedom for biquadratic NURBS (left) and bicubic NURBS (right).}
	\label{Ch6:fig:L2_sec41}
\end{figure}

For all cases, the patches are constructed using the same conforming mesh discretization. Two studies are conducted, employing different NURBS basis functions: biquadratic and bicubic. Note that in both studies, the functions maintain equal degrees in the two parameter directions of the patch. The results show a more pronounced convergence rate of relative error in the $L2$-norm for the bicubic NURBS functions than the convergence rate for the biquadratic NURBS functions, as depicted in Figures \ref{Ch6:fig:L2_sec41} on a unified log-log scale. The convergence rate of IGA solution using the Nitsche method shows consistency in approximating multiple patch domains and interfaces. The results obtained are in line with those from a single patch analysis.

\subsection{Rectangular piezoelectric energy harvester}

The purpose of this example is to investigate the effectiveness of multi-patch IGA in the context of PEH. To achieve this, we have chosen a widely studied piezoelectric harvester configuration \cite{erturk2008distributed, junior2009electromechanical, erturk2009experimentally}. It consists of two PZT-5A sheets connected in series and attached to a brass substructure, with a tip mass of $M_{tip}$=0.012kg. The geometric properties of this configuration are presented in Table \ref{Geo}.

\begin{figure}[ht]
	\centering
	\includegraphics[width=1.0\textwidth]{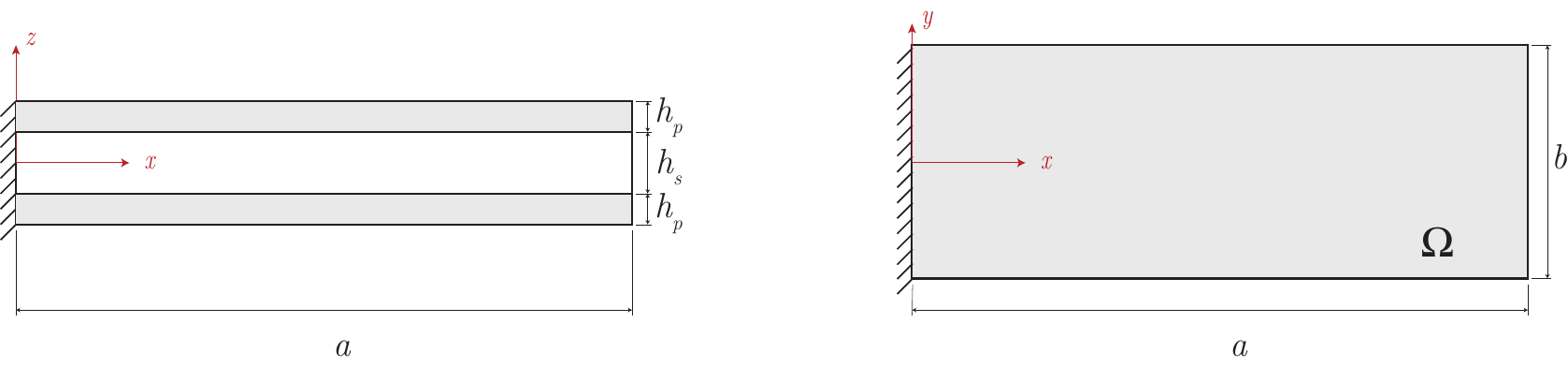}
\caption{Schematic of the bimorph piezoelectric energy harvester with a rectangular shape, consisting of two piezoelectric layers and one substructure layer. The $x-z$ plane is on the left, and the $x-y$ plane is on the right.}
	\label{Ch6:fig:schema}
\end{figure}

 \begin{table}[ht]
\centering
\caption{Geometric parameters of the studied PEH device.}
\renewcommand{\arraystretch}{1.25}
\label{Geo}
\begin{tabular}{@{}lrl}
\hline
Length $a$                              &50.8             & mm\\
Width $b$                               &31.8             & mm\\
Piezoelectric thickness $h_p$           &0.26             & mm\\
Substructure thickness  $h_s$           &0.14             & mm\\
\hline
\end{tabular}
\end{table}

\begin{figure}[ht]
	\centering
	\includegraphics[width=1.0\textwidth]{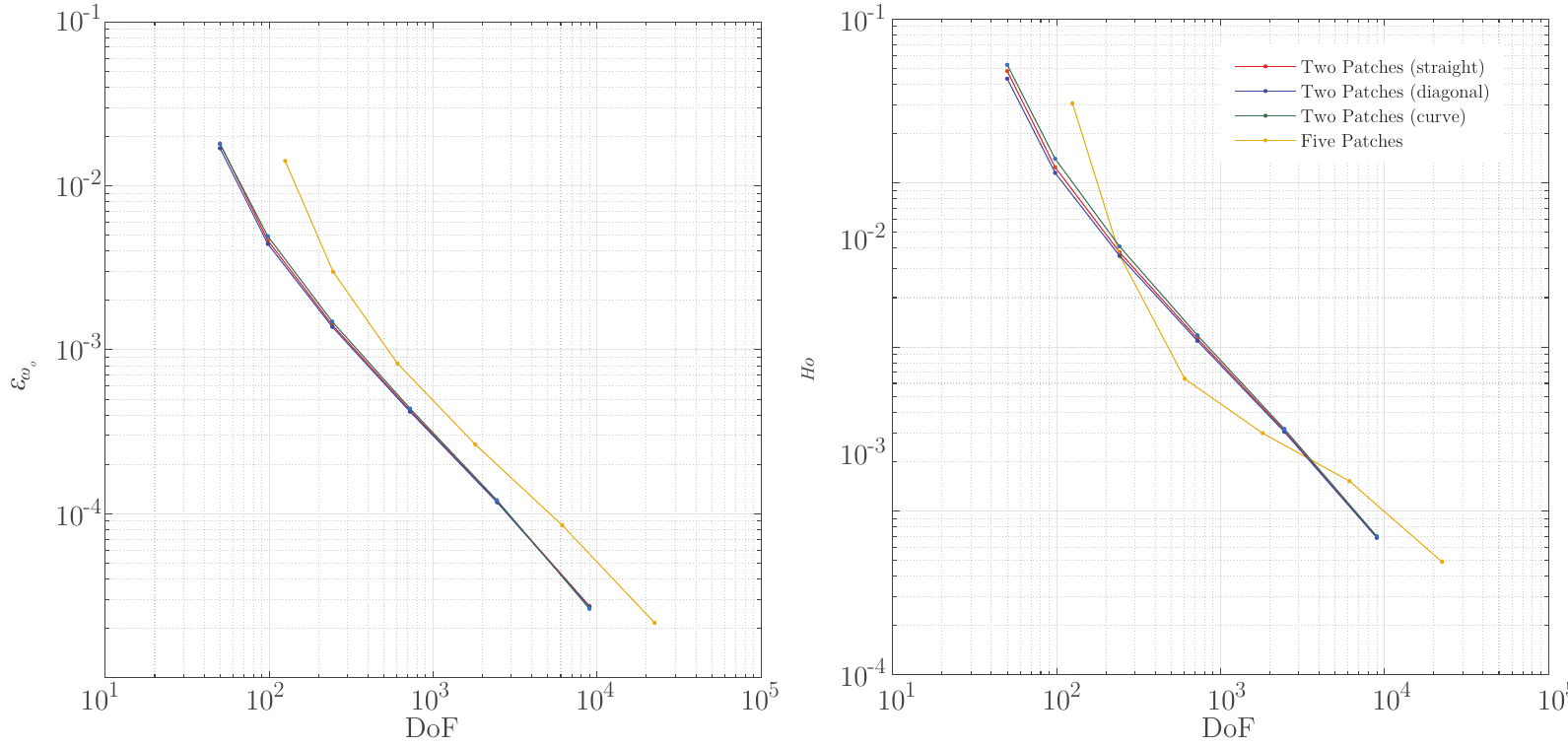}
\caption{Comparison of multi-patches IGA model in terms of $\varepsilon_{\omega_o}$ and $\varepsilon_{H_o}$ vs. the number of degrees of freedom.}
	\label{Ch6:fig:conv_wo_Ho}
\end{figure}

Similar to the previous example, the rectangular geometry of the piezoelectric device is constructed using multi-patches in accordance with the four cases outlined earlier and depicted in Figure \ref{Ch6:fig:caseA}. These four cases are evaluated in terms of accuracy for the fundamental frequency ($\omega_o$) and the ratio of voltage to acceleration ($H_o$) based on the number of degrees of freedom. The estimated $\omega_o$ and $H_o$ are then compared with the reference solution ($\omega_o^*$ and $H_o^*$) obtained using the single patch IGA model with a high-resolution mesh (67081 degrees of freedom). The parameters $\varepsilon_{\omega_o}$ and $\varepsilon_{H_o}$ are defined as

 \begin{equation}
     \varepsilon_{\omega_o} = \frac{\omega_o-\omega_o^*}{\omega_o^*};\ \varepsilon_{H_o} = \frac{H_o-H_o^*}{H_o^*};
 \end{equation}
The convergence rates of $\varepsilon_{\omega_o}$ and $\varepsilon_{H_o}$ are presented in Figure \ref{Ch6:fig:conv_wo_Ho}$a$ and \ref{Ch6:fig:conv_wo_Ho}$b$, respectively. The results demonstrate that the multi-patch IGA approach based on the Nitsche method delivers satisfactory convergence for PEH applications. 

Accurately modelling the geometry of complex shapes often involves a higher number of patches, expanding the system's degrees of freedom. As a result, the computational time needed to solve the system significantly increases. In a prior study, Peralta-Braz et al. \cite{peralta2023design} introduced a Model Order Reduction for an IGA-based model of PEH to alleviate the computational burden. In simpler terms, the deflection solution \textbf{w} is approximated by employing a truncated expansion of the first $K$ mode shape vectors corresponding to the first $K$ natural frequencies. Refer to Appendix A for further details.

\begin{figure}[ht]
	\centering
	\includegraphics[width=0.9\textwidth]{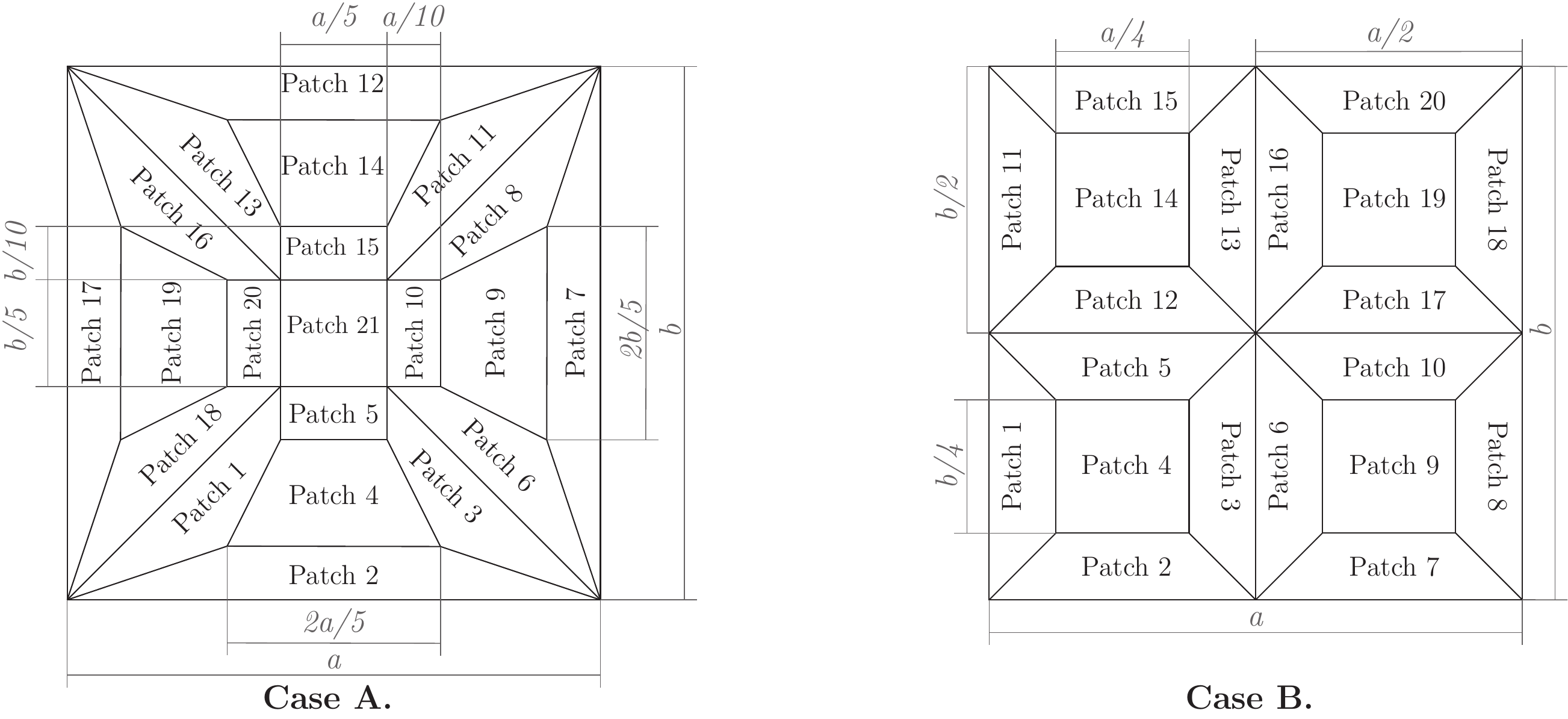}
\caption{Cases studied for the construction of the rectangular piezoelectric harvester.}
	\label{Ch6:fig:casePEH}
\end{figure}

\begin{figure}[ht]
	\centering
	\includegraphics[width=0.75\textwidth]{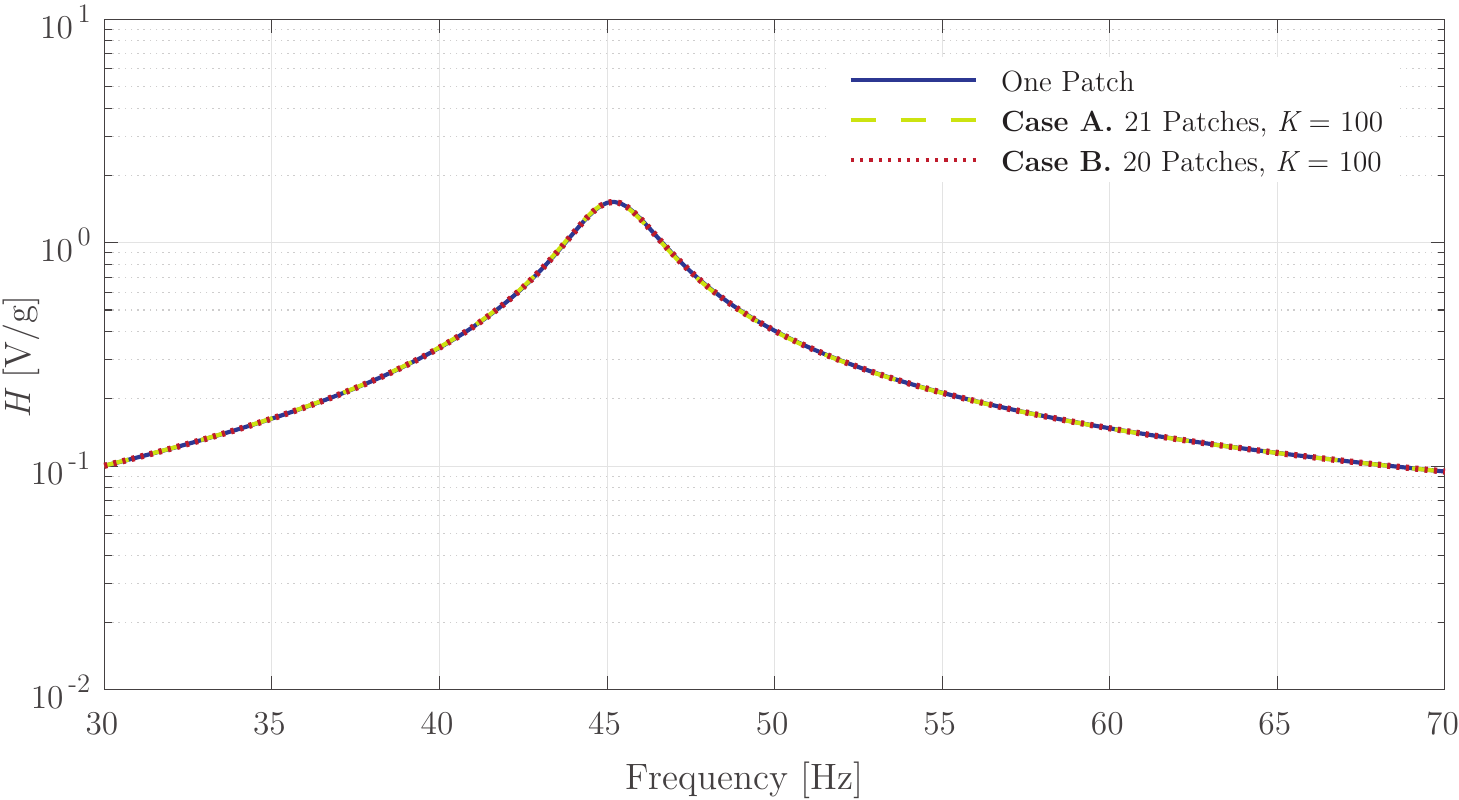}
        \caption{Comparison of Frequency Response Functions (FRFs) for a rectangular piezoelectric energy harvester (PEH) with different patch compositions, including a single patch, a composition of 21 patches (\textbf{Case A.}), and  a composition of 20 patches (\textbf{Case B.}). The cases of multiple patches considers a modal reduction considering the first 100 natural frequencies. }
	\label{Ch6:fig:FRFrectangular}
\end{figure}

Thus, two approaches for constructing the rectangular shape of piezoelectric devices are explored, employing modal order reduction to assess their performance. The approaches are presented in Figure \ref{Ch6:fig:casePEH}, where \textbf{Case A} is characterised by a complex conformation that utilises 21 patches while \textbf{Case B} adopts a simpler conformation employing 4 patches that are arranged in a 2$\times$2 grid pattern. In the modal reduction, 100 mode shape vectors are considered. These vectors are calculated using the built-in function \verb!eigs!, which is based on the Krylov-Schur iterative algorithm. Figure \ref{Ch6:fig:FRFrectangular} provides a comparison between the FRFs obtained from single-patch cases and those from \textbf{Case A} and \textbf{Case B}, using bicubic NURBS functions. These results illustrate the robustness of the method across a broad frequency range.

\subsection{Variable-Width piezoelectric energy harvester}
Recent research suggests that the utilisation of PEH with rectangular geometries does not yield optimal efficiency. In this direction, Rahimzadeh et al. \cite{rahimzadeh2021analysis} conducted an investigation on the output voltage of various PEH shapes and discovered that the variable-Width geometry had the most favourable performance among the configurations studied. This geometry showed an enhancement of 56\% in voltage output compared to rectangular geometries. Also, this example is noteworthy as it enables evaluating the Nitsche method in non-conforming scenarios, which requires careful consideration when designing and implementing numerical methods. Specifically, non-conforming scenarios are those where the meshes of patches along the boundary either do not align perfectly or exhibit different levels of discretization.

\begin{figure}[ht]
	\centering
	\includegraphics[width=1.0\textwidth]{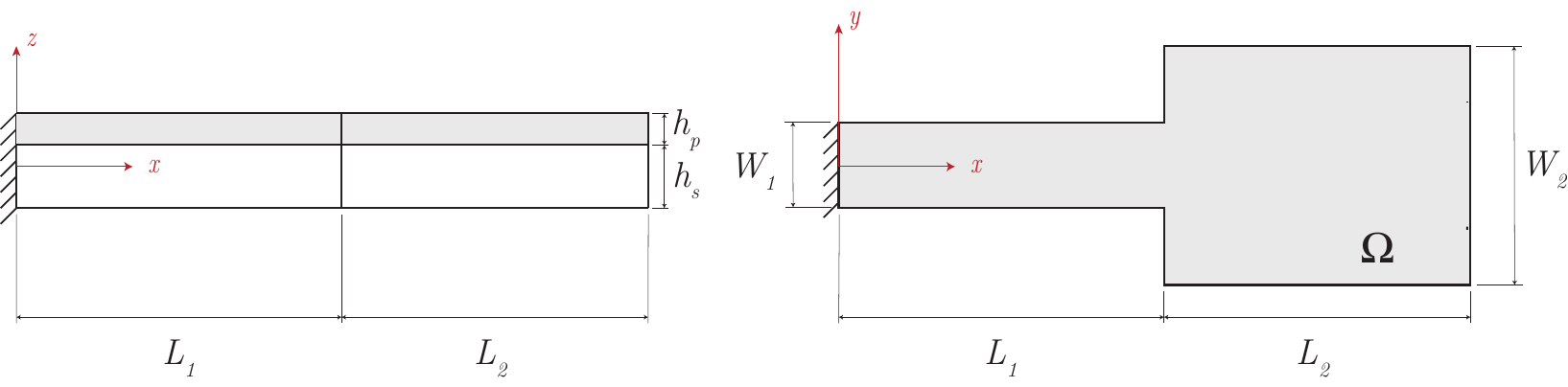}
\caption{Schematic of the unimorph piezoelectric energy harvester with a variable-width, consisting of one piezoelectric layer and one substructure layer.}
	\label{Ch6:fig:TwoStep}
\end{figure}

The design of the variable-width PEH is depicted in Figure \ref{Ch6:fig:TwoStep}, which is composed of a steel substructure with a thickness of 0.8mm and a PZT5H layer with a thickness of 0.4mm. The widths of the two steps, $W_1$ and $W_2$, are 12mm and 36mm, respectively. The total length of the beam is set at 75mm, while the lengths of each step are adjusted to maintain a constant overall length ($L_1 + L_2 = cte$). The narrower side of the plate is fixed by clamping.

\begin{figure}[ht]
	\centering
	\includegraphics[width=1.0\textwidth]{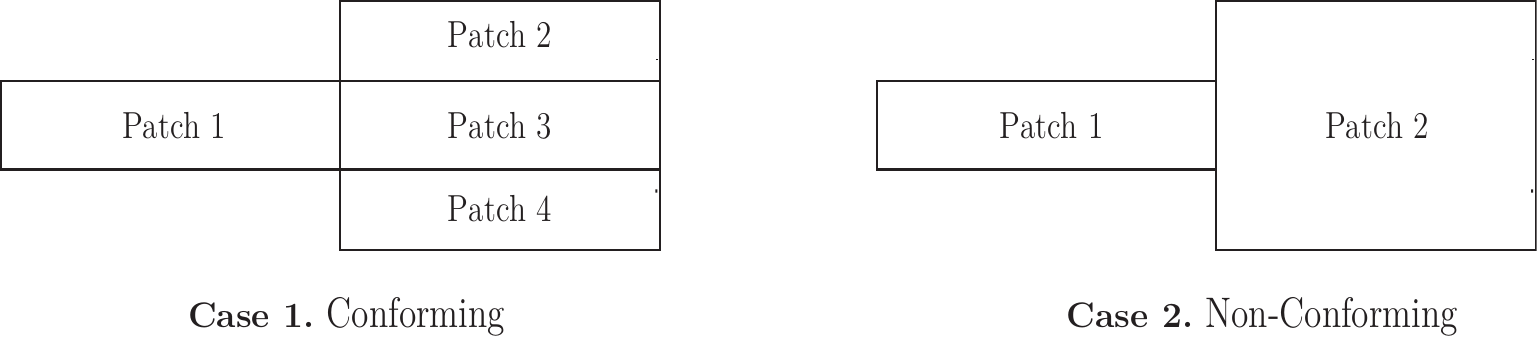}
\caption{Cases studied for the construction of the variable-width piezoelectric device.}
	\label{Ch6:fig:A_TwoStep}
\end{figure}

Two approaches for constructing the Two-Step PEH with multi-patches were evaluated, as shown in Figure \ref{Ch6:fig:A_TwoStep}. The first approach, \textbf{Case 1}, uses four NURBS patches with a consistent mesh discretisation, employing biquadratic functions. In contrast, the second approach, \textbf{Case 2}, employs two non-conforming patches with the exact discretisation using biquadratic functions.

The fundamental frequencies of the variable-width PEH calculated using the two approaches are presented along with the solutions from the literature in Table \ref{t:twoStep}. The relative difference between the IGA results and the reference solutions is given in parentheses. The results show good agreement with the literature, with differences of no more than 0.5\%.

\begin{table}[ht]
\centering
\caption{Geometric parameters of the studied PEH device and their estimated natural frequencies}
\renewcommand{\arraystretch}{1.25}
\label{t:twoStep}
\begin{tabular}{rrrrlrl}
\cline{3-7}
\multicolumn{1}{l}{}               & \multicolumn{1}{l}{}               & \multicolumn{5}{c}{Fundamental Frequency {[}Hz{]}}                                                                     \\ \hline
\multicolumn{1}{c}{$L_1$ {[}mm{]}} & \multicolumn{1}{c}{$L_2$ {[}mm{]}} & \multicolumn{1}{c}{Rahimzadeh et al.}  & \multicolumn{2}{c}{Conforming Case} & \multicolumn{2}{c}{Non-Conforming Case} \\ \hline
12.5                               & 62.5                               & 90.9                                   & \multicolumn{2}{r}{90.86 (0.04\%)}    & \multicolumn{2}{r}{90.88 (0.02\%)}        \\
25.0                               & 50.0                               & 82.4                                   & \multicolumn{2}{r}{81.12 (0.34\%)}    & \multicolumn{2}{r}{82.13 (0.32\%)}        \\
37.5                               & 37.5                               & 80.5                                   & \multicolumn{2}{r}{80.31 (0.23\%)}    & \multicolumn{2}{r}{80.32 (0.22\%)}        \\
50.0                               & 25.0                               & 83.4                                   & \multicolumn{2}{r}{83.47 (0.08\%)}    & \multicolumn{2}{r}{83.47 (0.08\%)}        \\
62.5                               & 12.5                               & 94.3                                   & \multicolumn{2}{r}{94.76 (0.48\%)}    & \multicolumn{2}{r}{94.76 (0.48\%)}        \\ \hline
\end{tabular}
\end{table}

\subsection{Piezoelectric energy harvester with a trapezoidal hole}

In line with the examples discussed in the previous section, this section focuses on a PEH with a trapezoidal hollow structure. Xu et al. \cite{xu2022design} investigated this problem with the goal of reducing its resonant frequency while simultaneously increasing the voltage output compared to rectangular devices. The results indicate that the trapezoidal hollow structure enhances the output voltage by 34\% while reducing the natural frequency by approximately 12\%. This reduction in natural frequency is desirable because it enables the energy harvester to operate more effectively at lower frequencies, commonly encountered in ambient vibrations, resulting in improved energy capture and conversion efficiency.

\begin{figure}[ht]
	\centering
	\includegraphics[width=1.0\textwidth]{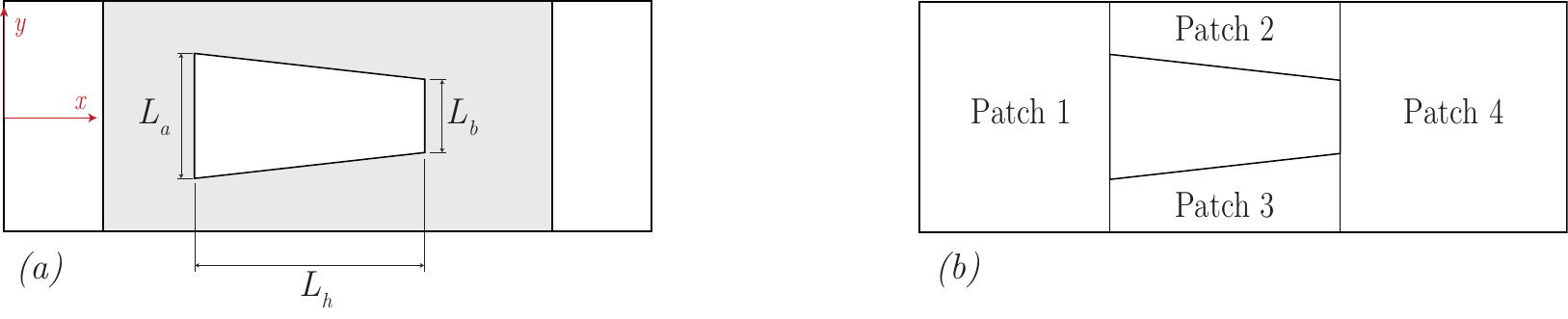}
\caption{$(a)$ Schematic of the unimorph piezoelectric energy harvester with a trapezoidal hole, consisting of one piezoelectric layer and one substructure layer. $(b)$ Patches used for the construction of the piezoelectric device with a trapazoidal hole.}
	\label{Ch6:fig:trapezoidalHole}
\end{figure}

Figure \ref{Ch6:fig:trapezoidalHole}$a$ presents the schema of the studied harvester. It consists of a copper substructure (80mm$\times$33mm$\times$0.2mm) and a PZT5H sheet (60mm$\times$33mm$\times$0.2mm). The trapezoidal hole has dimensions of $L_a$ = 20mm, $L_b$ = 15mm, and $L_h$ = 40mm. The geometry is constructed with four patches with a non-conforming mesh discretisation, as illustrated in Figure \ref{Ch6:fig:trapezoidalHole}$b$, utilising biquadratic NURBS basis functions. 

\begin{figure}[ht]
	\centering
	\includegraphics[width=0.75\textwidth]{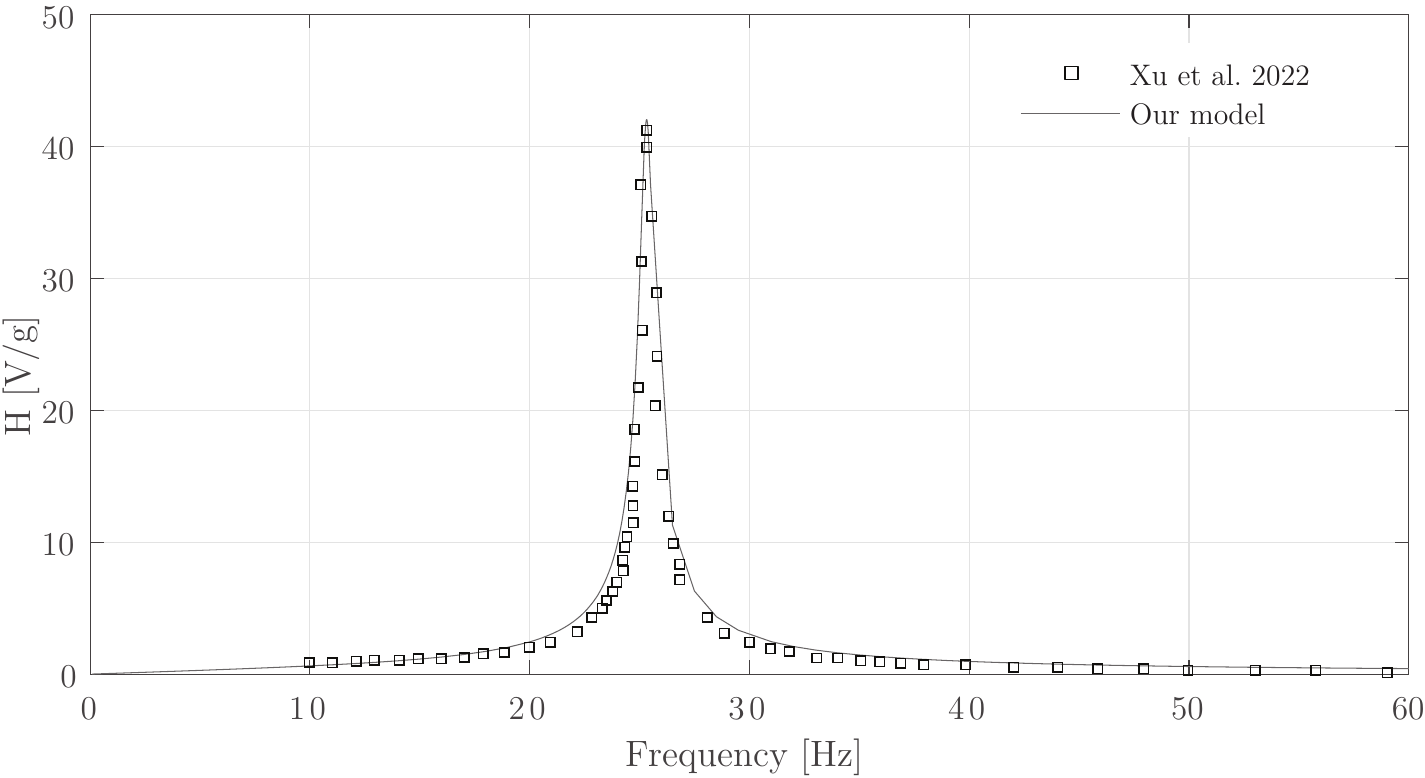}
\caption{Comparison of FRFs for the PEH with a trapezoidal hole employing the multi-patches IGA models and the results reported in reference \cite{xu2022design} 
}

	\label{Ch6:fig:FRFtrapezoidalHole}
\end{figure}

In Figure \ref{Ch6:fig:FRFtrapezoidalHole}, the voltages generated at different frequencies are compared with the results obtained by Xu et al. \cite{xu2022design} using a finite element approach. The grey solid lines represent the results of the multi-patches approach, while the red squares correspond to previous results reported in the literature \cite{xu2022design}. The comparison demonstrates an excellent agreement between the results, demonstrating the effectiveness of the proposed formulation in various contexts.

\section{Experimental validation using a metamaterial-inspired structure}
\label{S:5}

As discussed previously, the use of piezoelectric materials in combination with metamaterial structures is a promising technology for harvesting energy from ambient vibrations. Modelling the behaviour of such structures is crucial for understanding their performance and optimising their design. The multi-patch IGA model presented here is a valuable tool for this purpose, as it enables an accurate representation of the complex geometry and behaviour of the structure. In this section, we employ the presented model to simulate a PEH with a metamaterial-inspired structure, which was previously presented by Khattak et al. \cite{khattak2022concurrent}.

\begin{figure}[ht]
    \centering
    \includegraphics[width=0.5\textwidth]{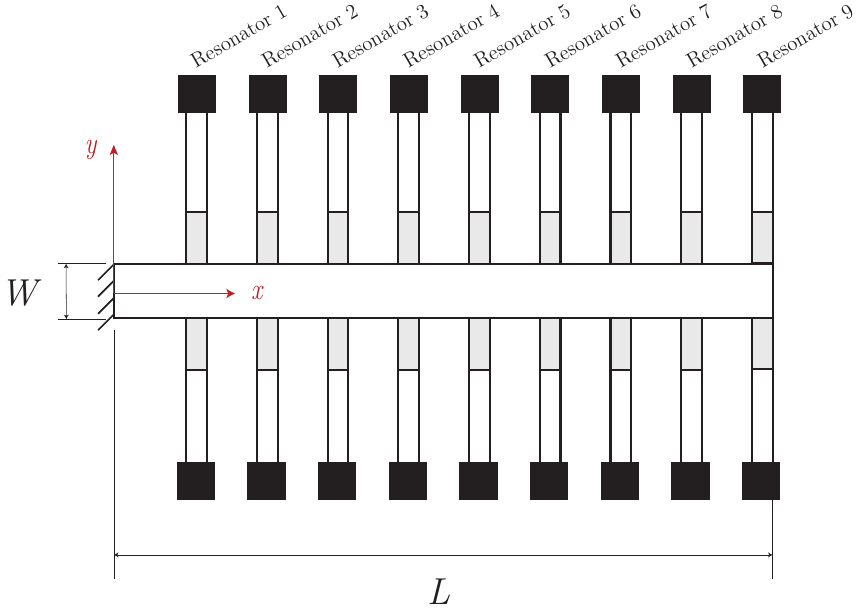}
    \caption{Schematic of the piezoelectric energy harvester with a meta-structure, consisting of eighteen piezoelectric resonators joint to the main plate.}
    \label{Ch6:fig:schemaMeta}
\end{figure}

\begin{figure}[ht]
	\centering
	\includegraphics[width=1\textwidth]{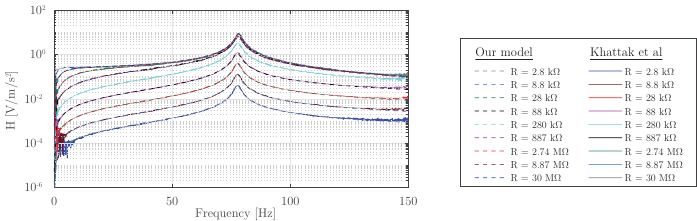}
\caption{Comparison of the FRF voltages of our resonator design to those used in \cite{khattak2022concurrent}, employing different resistors.}
	\label{Ch6:fig:14}
\end{figure}

The structure under study is a rectangular main structure with eighteen identical piezoelectric energy harvesters (PEHs) attached to it, as shown in Figure \ref{Ch6:fig:schemaMeta}. The main structure is composed of aluminium, with its parameters are summarised in Table \ref{t:MainStructure_beam}. However, the electromechanical parameter and the geometrical attributes of the resonator, used in \cite{khattak2022concurrent}, are not disclosed, but their equivalent lumped-element parameters are provided. Consequently, the bimorph piezoelectric resonators used in this validation are designed to match the damping coefficient ($c$), equivalent stiffness ($k$), electromechanical coupling ($\theta$), and capacitance ($C_p$) as outlined in the Lumped-Element model. Their explicit formulations are available in Appendix B for reference. The comparison between the parameters of our resonator design and those presented in \cite{khattak2022concurrent} is illustrated in Table \ref{t:resonators}. Furthermore, Figure \ref{Ch6:fig:14} compares the voltage frequency response function (FRF) of our piezoelectric resonator with the results obtained from an experimental characterisation of an isolated PEH conducted by Khattak et al., considering a range of different electrical resistances. The comparison demonstrates good agreement for the various values of electrical resistances.

\begin{table}[ht]
\centering
\caption{Parameters of the main structure of the meta-material PEH.}
\renewcommand{\arraystretch}{1.25}
\label{t:MainStructure_beam}
\begin{tabular}{llll}
\hline
Young's Modulus ($E$) & 70 GPa     &  Lenght ($L$)          & 0.9144 m \\
Poisson Modulus($\nu$)  & 0.3  & Width ($W$)           & 0.0254 m      \\
Density ($\rho$)     & 2700 kg/m$^3$ & Thickness ($t$)       & 1.49 mm   \\
Mass of clamps ($M_{clamps}$)      & 1.7 grs   & & \\ \hline
\end{tabular}
\end{table}

\begin{table}[ht]
\centering
\caption{Parameters of the piezoelectric resonator of the meta-material PEH.}
\renewcommand{\arraystretch}{1.25}
\label{t:resonators}
\resizebox{\textwidth}{!}{  
\begin{tabular}{lll|ll}
\hline
Parameter                          & Our work         & Khattak et al.   & Parameter                         & Our work  \\ \hline
Damping coefficient ($c$)            & 0.0525 N/m/s     & 0.0525 N/m/s     & Substructure´s Young modulus ($E$)  & 725 GP \\
Equivalent stiffness ($k$)           & 853.08 N/m       & 853.63 N/m       & Substructure's Poison modulus ($\nu$) & 0.3       \\
Electromechanical coupling ($\theta$) & -1.8707 10$^{-4}$ N/V & -1.8884 10$^{-4}$ N/V & Substructure's density ($\rho_s$)        & 7850 kg/m$^{3}$     \\
Capacitance ($C_p$)                       & 3.23 GF          & 3.23 GF          & Piezoelectric's density ($\rho_p$)          & 7800 kg/m$^{3}$      \\
Substructure's length ($L_s$)         & 0.029 m          &                  & Piezoelectric's ($c_{11}$, $c_{22}$)          & 69.2 GPa  \\
Piezoelectric's length ($L_p$)         & 0.010 m          &                  & Piezoelectric's ($c_{12}$)               & 24.3 GPa  \\
Width ($W$)                          & 0.003 m          &                  & Piezoelectric's ($c_{66}$)               & 22.6 GPa  \\
Substructure's thickness ($t_s$)      & 0.330 mm          &                  & Piezoelectric's ($e_{31}$, $e_{32}$)          & -16 c/m$^2$  \\
Piezoelectric's thickness ($t_p$)     & 0.045 mm          &                  & Piezoelectric's ($\varepsilon_{33}$)         & 9.57 nF/m \\ \hline
\end{tabular}
}
\end{table}

The relative displacement transmissibility at any given point, $\textbf{x}_{out}$, on the plate with respect to the base due to harmonic acceleration $a_b(t)=A_be^{i\omega t}$ is defined as follows:
\begin{equation}
    \label{Eq.TRANS}
    \text{TR} = \frac{w(\textbf{x}_{out}, \omega)-A_b/\omega^2}{A_b/\omega^2} 
\end{equation}

\begin{figure}[ht]
	\centering
	\includegraphics[width=0.75\textwidth]{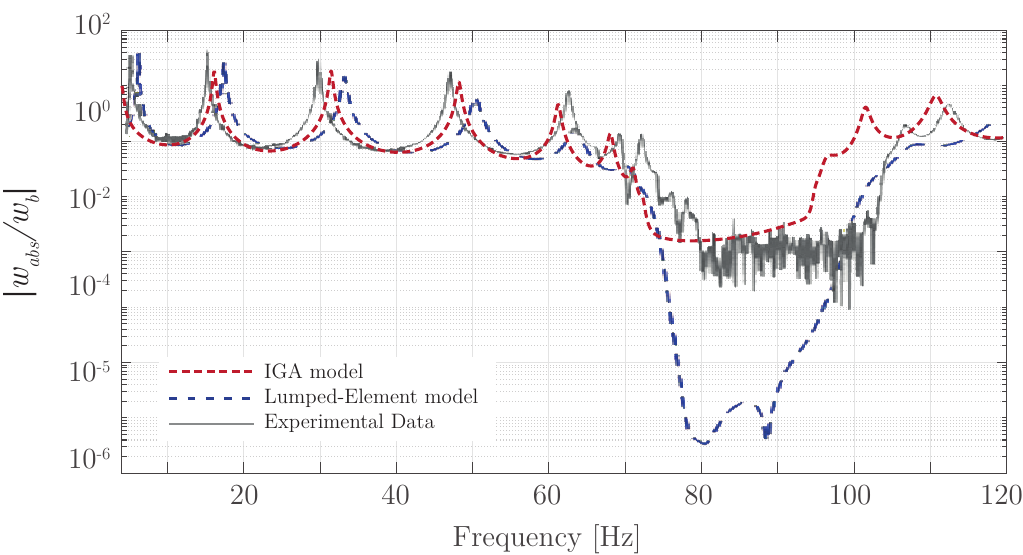}
\caption{Transmissibility comparison between experimental and analytical solutions from the literature \cite{khattak2022concurrent} and IGA model.}
	\label{Ch6:fig:MetaInspired}
\end{figure}

Figure \ref{Ch6:fig:MetaInspired} displays the transmissibility at the tip ($\textbf{x}_{out}$ = 0.9144m) obtained from the IGA model, which is compared to the theoretical and experimental data presented in reference \cite{khattak2022concurrent}. The results demonstrate that the integration of resonant energy harvesters can induce the formation of a locally resonant bandgap, as confirmed by experimental data. The IGA model shows a good agreement with the literature results. However, as noted by Khattak et al., discrepancies between the experimental results and the model can be attributed to variations in system properties and the influence of the clamping hardware of each resonator. This notion is further corroborated by the conclusions drawn by Peralta et al., who assert that variations in the electromechanical attributes and the mounting process of the PEH constitute the primary source of variability in the FRF. Similarly, discrepancies with the theoretical results could be due to the simplification of each energy harvester in the Lumped-Element model. In light of this, the Bayesian inference framework proposed by Peralta et al. in [23] could be employed to identify the current characteristics of the PEH and to select the most suitable model based on experimental observations.

\begin{figure}[ht]
	\centering
	\includegraphics[width=1\textwidth]{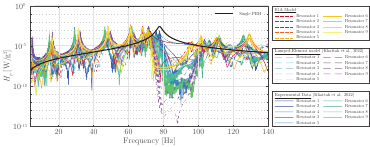}
\caption{Comparison of power FRF of each resonator and the single PEH.}
	\label{Ch6:fig:FRFresonators}
\end{figure}

Figure \ref{Ch6:fig:FRFresonators} shows the Power FRF of nine resonators identified in Figure \ref{Ch6:fig:schemaMeta}. This is contrasted with the Power FRF of a single piezoelectric resonator that is not adhered to the structure., but rather attached directly to the vibration source. The objective is to highlight the advantages of positioning one of these resonators directly at the source of vibration instead of incorporating it into the metastructure. Each resonator is connected to a load of 280 k$\Omega$. The maximum power is achieved at frequency values just before the bangap range, reaching power values higher than that of a single PEH. In addition, some piezoelectric resonators can produce more power compared to single PEH at low-frequency ranges. Consistent with Khattak et al. findings, the resonators located closer to the base of the main structure generate greater electrical power, indicating that the current configuration is suboptimal. These results motivate the need for an optimisation framework to maximise the power output of the structure.

\section{Piezoelectric shunt damping metamaterial for vibration attenuation}
\label{S:6}
\begin{figure}[h!]
    \centering
    \includegraphics[width=0.85\textwidth]{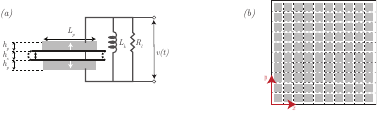}
    \caption{$(a)$ Schematic representation of the bimorph single cell. Piezoelectric sheets are interconnected in series, while the inductance and electrical resistance are connected in parallel. $(b)$ Schematic of the piezoelectric metamaterial, featuring a grid of 10$times$10 cells. The dotted lines delineate the boundaries between individual cells.}
    \label{Ch6:fig:scheme_Agh}
\end{figure}

As discussed previously, piezoelectric materials based on the shunt damping approach have shown promise in creating bandgaps \cite{aghakhani2020modal,sugino2020analytical}. This serves as an alternative to the use of mechanical resonators, which could be beneficial in contexts where the addition of mass or volume occupancy may present challenges. Here, piezoelectric sheets are integrated into vibrating host structures and connected to a resonant shunt circuit. This circuit consists of the piezoelectric sheet connected in parallel with an inductive element $L_k$ and an electrical resistance $R_l$, as illustrated in Figure \ref{Ch6:fig:scheme_Agh}a. The inductance value is chosen according to the target bandgap frequency $\omega_t$ as $L_k = (C_p\cdot \omega_t)^{-1}$, while larger resistance values favour attenuation. It's important to note that the electrical Equation \ref{eq:ELEC} is not valid in this case and should be updated as, 
\begin{equation}
    C_p \dot{v_p} + \left( \frac{1}{R_l} + \frac{1}{jL_k} \right) v_p + \boldsymbol{\Theta}^{T}\textbf{w} =0 
\label{eq:ELEC_new}
\end{equation}

Aghakhani et al. \cite{aghakhani2020modal} delved into studying this type of piezoelectric metamaterial, presenting a comprehensive system-level modal analysis framework. Their work showcased the efficacy of generating a bandgap at a target frequency for effective vibration suppression. To validate our current model, we replicate a numerical example presented by Aghakhani et al. \cite{aghakhani2020modal}. The configuration consists of an aluminium host plate measuring 800mm $\times$ 700mm $\times$ 1mm, with PZT-5H sheets of 1mm thickness attached on both sides. The plate is fully clamped on all four sides. A grid of 10$\times$10 independent and identical electrodes is considered, with each modelled as a patch, as shown in Figure \ref{Ch6:fig:scheme_Agh}b. The inductance is selected considering a target bandgap frequency $\omega_t$ of 300 Hz, while the chosen electrical resistance is 1 M$\Omega$. A harmonic point force is applied at the point $\textbf{x}_{in}$ = (680mm, 588mm), and it oscillates with a frequency denoted as $\omega$. The transmissibility deformation resulting from this force can be defined at any arbitrary point $\textbf{x}_{out}$ within the plate. The  expression for this transmissibility is defined as follows,
\begin{equation}
    \label{Eq.wh}
    \text{TR} = \frac{w(\textbf{x}_{out}, \omega)}{w(\textbf{x}_{in}, \omega)} 
\end{equation}

\begin{figure}[h!]
    \centering
    \includegraphics[width=0.85\textwidth]{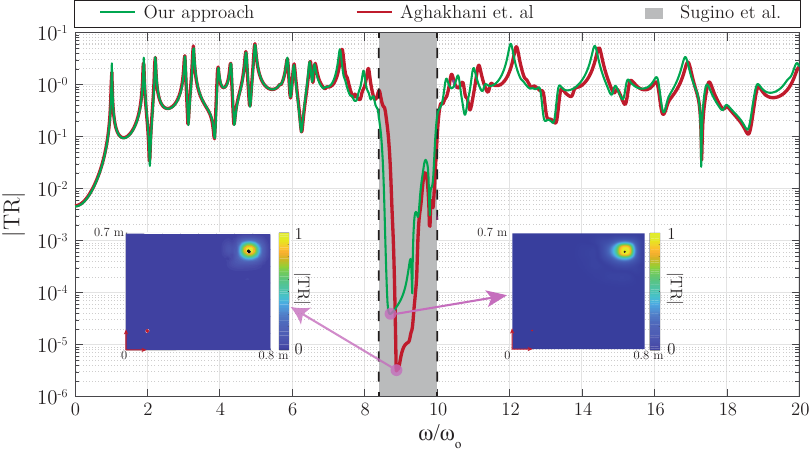}
    \caption{Comparison of transmissibility frequency response functions between the forcing point and sensing point $\textbf{x}_{out}^{*}$ = (120mm, 112mm) utilising both the system-level modal analysis framework and the multi-patches IGA framework. The inset plots illustrate the transmissibility amplitude across the plate at the frequency that minimizes the transmissibility at the sensing point $\textbf{x}_{out}^{*}$}
    \label{Ch6:fig:Trans_Agh}
\end{figure}

Figure \ref{Ch6:fig:Trans_Agh} presents the transmissibility function at point $\textbf{x}_{out}^{*}$ = (120mm, 112mm) estimate from the present model, contrasting it with the results presented in reference \cite{aghakhani2020modal}. Additionally, the results are compared with the bandgap analytical predictions based on the work by Sugino et al. \cite{sugino2020analytical}, as also done in reference \cite{aghakhani2020modal}. Furthermore, Figure \ref{Ch6:fig:Trans_Agh} presents the transmissibility across the entire plate at the frequency $\omega^{*}$, which corresponds to the minimum value of the transmissibility function at the sensing point $\textbf{x}_{out}^{*}$ for both approaches. These minimum values are clearly marked and identified on the graph. Consequently, it is evident that vibration suppression extends beyond the sensing point and is effectively achieved across a significant portion of the entire plate.

In general, the estimated results demonstrate a strong alignment with existing literature, effectively identifying the bandgap. This illustrative example emphasises the versatility of the model in the field of piezoelectric materials. Moreover, it opens the discussion on the potential advantages of incorporating non-conventional electrode shapes on piezoelectric sheets, particularly considering the inherent benefits of IGA in providing accurate representations of complex geometries. Such considerations hold promise for enhancing the practical applications of piezoelectric metamaterials, opening new avenues for optimising their performance, whether as effective vibration suppressors or power generators.

\section{Plate metastructure with internal resonators}
\label{S:7}
\begin{figure}[ht]
    \centering
    \includegraphics[width=1\textwidth]{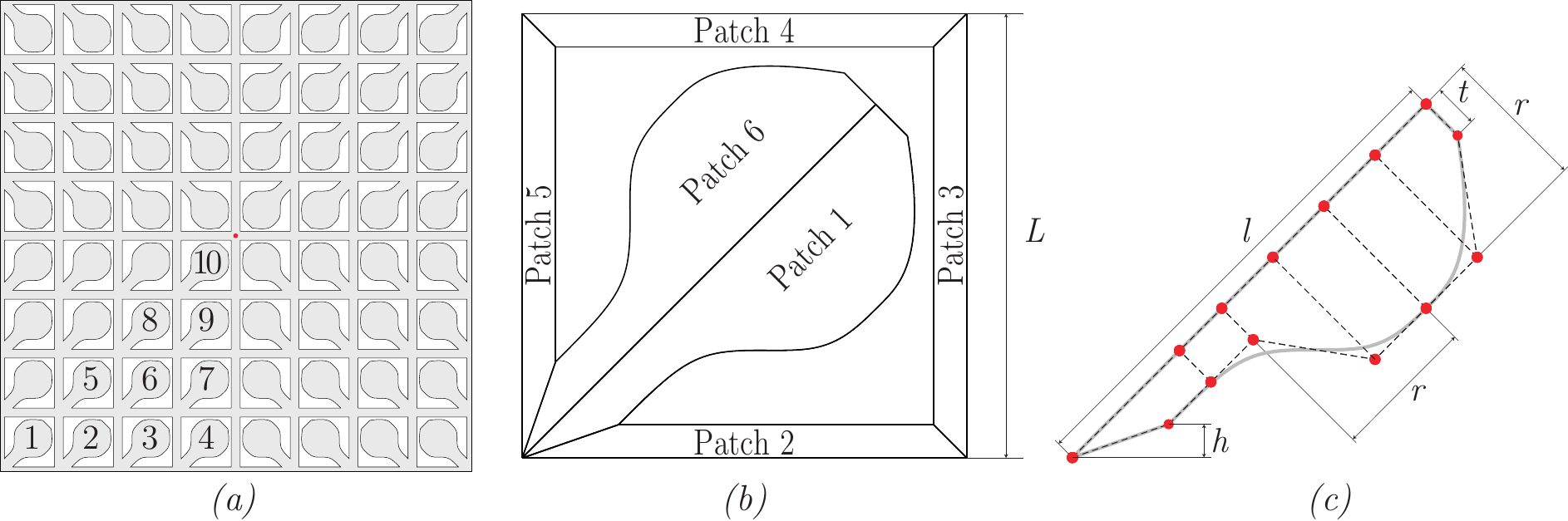}
    \caption{$(a)$ Schematic of patches employed in constructing the single cell within the plate metamaterial. $(b)$ Schematic of the piezoelectric plate metamaterial with internal resonator, featuring a grid of 8$\times$8 cells. The 10 key resonators are enumerated. The sensing point is denoted by a red dot in the centre. $(c)$ 2D NURBS surface representation of Patch 2 within a single cell, characterized by knot vectors $\Xi$ = \{0, 0, 0, 0, 1/4, 2/4, 3/4, 1, 1, 1, 1\} and $H$ = \{0, 0, 1, 1\}. The control points are marked by red circles. Patch boundaries are delineated by black lines. }
    \label{Ch6:fig:schema_Meta_Plate}
\end{figure}

In what follows, we present a novel meta-structure plate design. The design is inspired by previous studies in the literature, which show that meta-structures with internal resonators having mass-on-a-beam effects exhibit band-gap phenomenon. Taking advantage of the IGA parametrisation, we can explore various geometries of resonators, which represent various distributions of mass. These resonators' geometries can be finely tuned to optimise both energy harvesting and vibration suppression. Figure \ref{Ch6:fig:schema_Meta_Plate} depicts the design of the metastructure. Figure \ref{Ch6:fig:schema_Meta_Plate}a presents the 8$\times$8 metastructure grid. In Figure \ref{Ch6:fig:schema_Meta_Plate}b, the square unit-cell, composed of six separate patches, is illustrated. The corresponding control points of patch 1 and its parametrisation are shown in Figure \ref{Ch6:fig:schema_Meta_Plate}c. The parameterisation is done in terms of four parameters, $l$, $r$, $h$ and $t$, Whose values used in this implementation are presented in Table \ref{t:MainStructure}.

The metamaterial plate consists of a 1 mm aluminium structure with 1 mm layers of PZT-5A adhered to each side, which are connected in series. Each cell has an independent electrode, which is connected to an electrical resistance of 100 $\Omega$. The boundary conditions on the four edges are simple supports. The natural frequency $\omega_o$ of the meta-material is 48 Hz. The transmissibility in the centre of the plate $\textbf{x}_{out}^{*}$ = (200mm, 200mm) is studied. The formulation of the transmissibility FRF was defined in Equation \ref{Eq.TRANS}.

\begin{table}[ht]
\centering
\caption{Parameters of the main structure of the meta-material PEH.}
\renewcommand{\arraystretch}{1.25}
\label{t:MainStructure}
\begin{tabular}{ccccc}
\hline
$L$                       & $l$                         & $r$                         & $t$                      & $h$                        \\
\hline
\multicolumn{1}{r}{50 mm} & \multicolumn{1}{r}{56.3 mm} & \multicolumn{1}{r}{16.3 mm} & \multicolumn{1}{r}{5 mm} & \multicolumn{1}{r}{3.7 mm} \\ \hline
\end{tabular}
\end{table}

\begin{figure}[ht]
    \centering
    \includegraphics[width=0.75\textwidth]{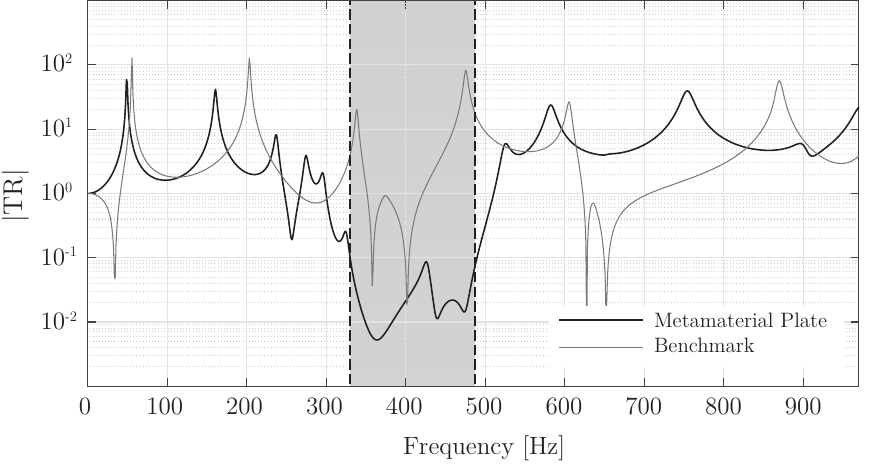}
    \caption{Transmissibility frequency response functions at the centre of the plate. The frequency is normalised with respect to the resonance frequency of the meta-material $\omega_o$ = 48 Hz.}
    \label{Ch6:fig:Trans_Meta_Plate}
\end{figure}

The FRF transmissibility at the centre of the plate with respect to the base support displacement is presented in Figure \ref{Ch6:fig:Trans_Meta_Plate}. It demonstrates the formation of a locally resonant bandgap in the system. This phenomenon is evidenced by a significant reduction in vibrations within the frequency range of approximately 320 to 480 Hz where the magnitude of the transmissibility reaches values lower than 10$^{-1}$. This reduction, within the locally resonant bandgap, is a consequence of the system’s capacity to efficiently suppress and dampen vibrations within a certain frequency spectrum. To qualify the effectiveness of vibration suppression, the FRF transmissibility is compared with that of a continuous plate, referred to as the Benchmark. This reference plate shares the meta-structure’s overall dimensions while its material composition and thickness are the same as its substructure. In Figure \ref{Ch6:fig:Trans_Meta_Plate}, it is evident that the benchmark plate attenuates vibrations within the bandgap frequencies. However, this effect is limited to specific frequencies. Even a slight deviation in the excitation frequency of the vibration source results in a substantial increase in plate displacement. In contrast, the metamaterial-based structure ensures vibration control within a 150 Hz range, as highlighted in the gray area in Figure \ref{Ch6:fig:Trans_Meta_Plate}. The innovative design of such structures holds promise for the design of the internal resonators to enhance suppression both in magnitude and frequency range, as well as to achieve a target bandgap frequency.

\begin{figure}[ht]
    \centering
    \includegraphics[width=0.9\textwidth]{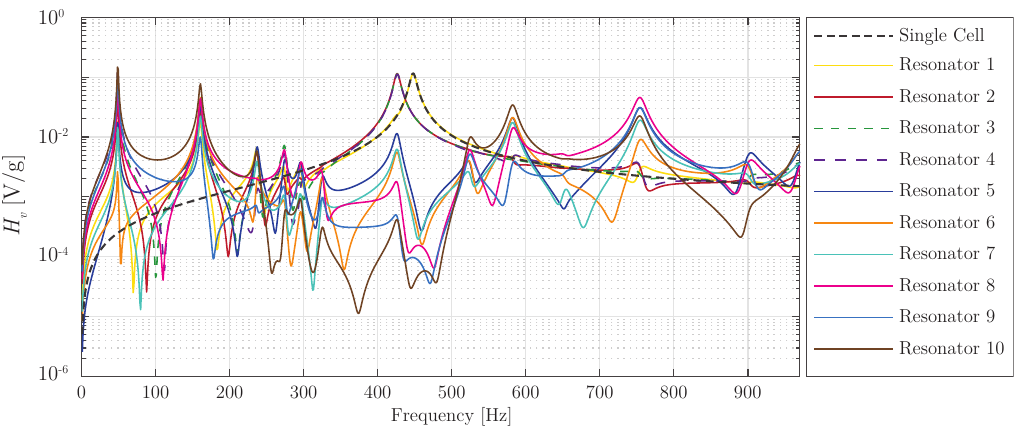}
    \caption{Comparison of Power Frequency Response Functions for the 10-Key Cells and the single cell.}
    \label{Ch6:fig:PEH_Meta_Plate}
\end{figure}

Figure \ref{Ch6:fig:PEH_Meta_Plate} presents the voltage Frequency Response Function (FRF) of ten key resonators, which are highlighted in Figure \ref{Ch6:fig:schema_Meta_Plate}. Given the symmetry of the structure, characterising these ten resonators is sufficient to understand the electro-dynamic behaviour of the remaining ones.  Analysing the voltage FRF within the range of 0 to 100 Hz, reveals that all piezoelectric resonators can generate more power than a single resonator attached to the same base vibration, with the resonance in the metastructure being eight times lower than that of the single resonator. In particular, a remarkable performance is seen in resonator 10, which is centrally located in the metamaterial array. In the range $[0-100]$ Hz, resonator 10 produces 350 times more voltage compared to a single cell. At 48 Hz, the voltage generation increases between 12 times (for resonator 6) and 760 times (for resonator 10). This indicates that resonators closer to the structure's centre exhibit superior performance at lower frequencies.

Focusing on the bandgap frequency range (443 $\pm$ 30 Hz), resonators 1, 2, 3, and 4, positioned near the excited base on the side of the structure, generate significantly more power than those farther from the excitation, closer to the center. They retain a voltage generation performance comparable to a simple generator when operating around these frequencies. In contrast, at lower frequencies, the resonators near the center demonstrate the best performance, emphasizing the critical importance of the resonator’s position within the structure. This position is intrinsically tied to the target frequency dictated by the application. As a result, certain resonators can be considered less effective depending on the target frequency. For instance, if the goal is to harvest energy at the bandgap frequencies, it would be prudent to focus only on positions 1, 2, 3, and 4, as their energy generation performance is significantly superior under such conditions.

\begin{figure}[ht]
    \centering
    \includegraphics[width=1\textwidth]{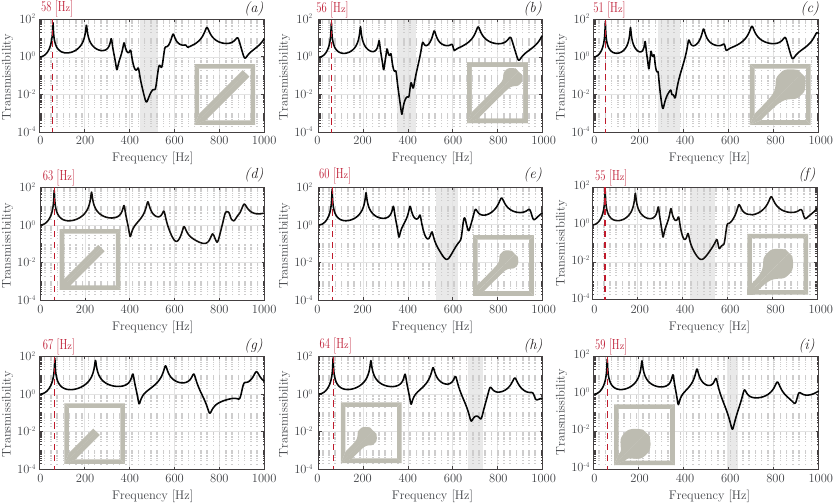}
    \caption{Comparison of transmissibility frequency response functions for nine metastructures with different internal resonator dimensions, which dimensions are: $(a)$ $r$ = 3.75 mm and $l$ = 37.5 mm, $(b)$ $r$ = 8.13 mm and $l$ = 37.5 mm, $(c)$ $r$ = 12.5 mm and $l$ = 37.5 m, $(d)$ $r$ = 3.75 mm and $l$ = 48.75 mm, $(e)$ $r$ = 8.13 mm and $l$ = 48.75 mm, $(f)$ $r$ = 12.5 mm and $l$ = 48.75 m, $(g)$ $r$ = 3.75 mm and $l$ = 60 mm, $(h)$ $r$ = 8.13 mm and $l$ = 60 mm, $(i)$ $r$ = 12.5 mm and $l$ = 60 mm. Their natural frequencies $\omega$ are denoted. The shape of the internal resonator is presented.}
    \label{Ch6:fig:PEH_Meta_Plate_ParaStudy}
\end{figure}

\begin{figure}[ht]
    \centering
    \includegraphics[width=0.5\textwidth]{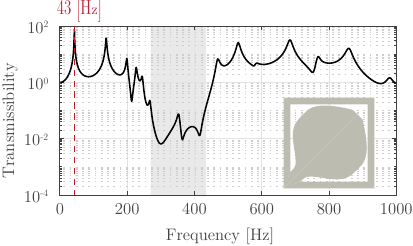}
    \caption{Transmissibility frequency response function
of a metastructures with $r$ = 21.25 mm and $l$ = 57.5 mm. The natural frequencies $\omega$ are denoted. The shape of the internal resonator is presented.}
    \label{Ch6:fig:PEH_Meta_Plate_ParaStudy2}
\end{figure}

Given that the resonators' design also influences the structure's dual functionality, a parametric study is conducted to explore this relationship. The parameters $r$ and $l$ control the diameter of the resonator tip and its length, respectively, which primarily determine the internal resonator's frequency. To further understand how transmissibility varies with these local resonator parameters, Figure \ref{Ch6:fig:PEH_Meta_Plate_ParaStudy} illustrates the transmissibility response for a series of parameter combinations. As $r$ increases, the initial frequency of the bandgap shifts to lower frequencies, and the bandgap width increases. Additionally, the natural frequency decreases with increasing $r$. Conversely, increasing $l$ shifts the bandgap to lower frequencies but narrows it. Notably, for small values of $r$, no bandgap occurs, which is especially unfavourable from the perspective of the vibration suppression. Meanwhile, the natural frequency increases proportionally to $l$. Based on this, high values of $r$ and $l$ result in a low frequency and wide bandwidth, as shown in Figure \ref{Ch6:fig:PEH_Meta_Plate_ParaStudy2}. Therefore, by adjusting the local resonator parameters, it is possible to control the formation of bandgaps and tune the resonance frequency of the metastructure, enabling efficient energy harvesting at low frequencies.

\section{Conclusions}
\label{S:8}
This work presents a numerical framework for piezoelectric energy harvesting (PEH) that employs a multi-patch isogeometric analysis procedure combined with Nitsche's method to discretize the governing equations and enforce continuity conditions between patches. The model allows the resonators to operate either as a unified energy harvester connected to a single electrical circuit or as a series of independent energy harvesters, each with its own circuit. This dual capability enhances both the system's flexibility and efficiency. The model's accuracy was validated against data from the literature for several benchmark configurations, demonstrating good agreement with the results.

In addition, we proposed a parameterized metastructure plate design featuring integrated resonators. Our research, supported by existing literature, demonstrates that positioning cells or resonators near the excitation source in the primary host structure significantly enhances power output within the bandgap range. Conversely, at low frequencies, resonators located near the center can amplify power by up to 760 times at the resonance frequency compared to a single resonator. Thus, depending on the application, certain resonator positions prove more efficient than others. Additionally, in terms of vibration attenuation, the design establishes a bandgap, effectively controlling vibration amplitudes within a 150 Hz range. Our findings suggest that the metastructure's performance in both energy harvesting and vibration damping can be further optimized.The proposed isogeometric analysis (IGA) model, designed for seamless integration with optimization algorithms, will be a key focus of future research. Another promising direction to explore is the development of graded metastructures, which show significant potential for improving both energy harvesting and vibration attenuation capabilities.

\section*{Acknowledgements}
This research is undertaken with the assistance of resources and services from the National Computational Infrastructure (NCI), which is supported by the Australian Government. The authors acknowledge support from the University of New South Wales (UNSW) Resource Allocation Scheme managed by Research Technology Services at UNSW Sydney.

\newpage
\appendix
\section{Non-Uniform Rational B-Splines (NURBS)}

B-Splines are polynomials that are defined piecewise on a knot vector. The knot vector is a non-decreasing sequence of $n+p+1$ real numbers $\xi_i$ (knots), given by:
\begin{equation}
\Sigma ={\xi_1 = 0,\xi_2,\dots,\xi_{n+p+1} = 1},
\end{equation}
where $n$ is the number of basis functions, and $p$ is the degree. B-Spline basis functions are recursively defined as follows:

\begin{equation}
\label{Ni0}
N_i^0(\xi)=\left\{
\begin{array}{cl}
1 \qquad & \text{if } \xi_i \leq \xi < \xi_{i+1} \\
 0 \qquad & \text{otherwise}
\end{array}
\right.
\end{equation}
and

\begin{equation}
\label{Nip}
N_i^p(\xi) = \frac{\xi-\xi_i}{\xi_{i+p}-\xi_i}N_i^{p-1}(\xi)+\frac{\xi_{i+p+1}-\xi}{\xi_{i+p+1}-\xi_{i+1}}N_{i+1}^{p-1}(\xi), \qquad p=1,2,3,\dots
\end{equation}

Non-Uniform Rational B-Splines (NURBS) are defined as:

\begin{equation}
\label{Rip}
R_i^p(\xi)=\frac{N_i^p(\xi)\bar{w}i}{\sum{j=1}^nN_j^p(\xi)\bar{w}_j},
\end{equation}
To avoid ambiguity with the plate deflection, the weights are denoted with a bar. If the weights are all equal, NURBS become B-Splines. To define a NURBS curve, a collection of NURBS basis functions and a set of $n$ control points $\textbf{P}_i$ are used:

\begin{equation}
\mathbf{C}(\xi) = \sum\limits_{i = 1}^n R_i^p(\xi)\mathbf{P}_i, \qquad \xi \in[0,1]
\end{equation}

A two-dimensional NURBS basis is defined by the tensor product of two one-dimensional NURBS bases, i.e.,

\begin{equation}
R_{i,j}^{p,q}(\xi,\eta)= \frac{N_{i}^{p}(\xi)N_{i}^{q}(\eta)\bar{w}{i,j}}{\sum{\hat{i}=1}^n\sum_{\hat{j}=1}^m N_{\hat{i}}^{p}(\xi)N_{\hat{j}}^{q}(\eta)\bar{w}_{\hat{i},\hat{j}}}
\end{equation}
where $N_j^q(\eta)$ are NURBS of degree $q$ defined on knot vector $\Pi = {\eta_0 = 0, \eta_1, \dots, \eta_{m + q + 1} = 1}$. A two-dimensional NURBS surface is then defined as:
\begin{equation}
    \label{S}
    \textbf{S}(\xi,\eta)=
    \sum_{i=1}^n\sum_{j=1}^m R_{i,j}^{p,q}(\xi,\eta)\textbf{P}_{i,j} \qquad (\xi, \eta)\in[0,1]\times[0,1]
\end{equation}
  where $\textbf{P}_{i,j}$ is a set of $n\times m$ control points with weights $\bar{w}_{i,j}$. 
  
  In this work,  following notation is employed:
\begin{equation}
    \label{S2}
    N_I(\boldsymbol{\xi}) = R_{i,j}^{p,q}(\xi,\eta), 
\end{equation}  
where $\boldsymbol{\xi}=(\xi,\eta)$ defines the parametric coordinate ($\boldsymbol{\xi}\in[0,1]\times[0,1]$) and $I = (i,j)$ is a two-dimensional index ($I = 1..k, k=n\times m$). Eq.(\ref{S}) defines a map between physical space $\mathbf{x} = (x,y)\in\Omega$ and parameter space $\boldsymbol{\xi}\in[0,1]\times[0,1]$:
\begin{equation}
    \label{x}
    \textbf{x}(\boldsymbol{\xi})=\sum^{k}_{I=1}N_I(\boldsymbol{\xi})\tilde{\textbf{x}}_I,
\end{equation}
where $\tilde{\textbf{x}}_I=\textbf{P}_{i,j}$. 
%%%%%%%%%%%%%%%%%%%%%%%%%%%%%%%%%%%%%%%%%%%%%%%%%%%%%%
%%%%%%%%%%%%%%%%%%%%%% Appendix B %%%%%%%%%%%%%%%%%%%%
%%%%%%%%%%%%%%%%%%%%%%%%%%%%%%%%%%%%%%%%%%%%%%%%%%%%%%
\section{Modal Order Reduction (MOR) }

Modal Order Reduction (MOR) aims to decrease the system's degrees of freedom by focusing on a restricted set of vibration modes to characterize dynamic behaviour within a specified frequency range. These modes are determined through mechanical equations under the assumptions of an unforced, undamped, and unpolarised device., i.e.

\begin{equation}
\label{red-eq}
    \textbf{M}\ddot{\textbf{w}} + \textbf{K}\textbf{w} = \textbf{0}
\end{equation}
Subsequently, expressing the deflection in a harmonic form gives rise to the generalised eigenvalue problem.
\begin{equation}
\label{eigenvalues-problem}
    (\textbf{K} - \omega_i^2 \textbf{M})\boldsymbol{\phi}_i = \textbf{0}
\end{equation}
$\boldsymbol{\phi}_i\in\mathbb{R}^{N\times1}$ represents the mode shape vectors, and $\omega_i$ signifies the undamped natural frequencies. The deflection solution \textbf{w} can be approximated through a truncated expansion of the first $K$ mode shape vectors, associated with the first $K$ natural frequencies as

\begin{equation}
\label{deflection-approximation}
    \textbf{w}\approx\textbf{w}_o = \boldsymbol{\Phi}_o \boldsymbol{\eta}
\end{equation}
where $\textbf{w}_o\in\mathbb{R}^{N \times 1}$ is the approximate deflection solution, $\boldsymbol{\Phi}_o\in\mathbb{R}^{N \times K}$ is the matrix which contains the first $K$ mode shape vectors $\boldsymbol{\phi}_i$ and $\boldsymbol{\eta}\in\mathbb{R}^{K\times1}$ denotes the modal coordinates. In this way, it is possible to define the reduced mass matrix $\textbf{m}_o$:
\begin{equation}
\label{reduced-mass}
    \textbf{m}_o = \boldsymbol{\Phi}_o^T \textbf{M} \boldsymbol{\Phi}_o
\end{equation}

Next, equations (\ref{eq:mech_02}) and (\ref{eq:elec_02}) can be represented in modal coordinates and normalised by the reduced mass matrix resulting in a reduced-order system
\begin{equation}
\label{SistEq-1R}
    \ddot{\boldsymbol{\eta}}+\textbf{c}_o\dot{\boldsymbol{\eta}}+ \textbf{k}_o\boldsymbol{\eta} - \boldsymbol{\theta}_ov_p = \textbf{f}_o a_b
\end{equation}

\begin{equation}
\label{SistEq-2R}
    C_p\dot{v}_p + \frac{v_p}{R_l} + \boldsymbol{\Theta}^T\boldsymbol{\Phi}_o \dot{\boldsymbol{\eta}}=0 
\end{equation}
where

\begin{equation}
    \begin{split}
       \textbf{c}_o & = \textbf{m}_o^{-1} \boldsymbol{\Phi}_o^T \textbf{C} \boldsymbol{\Phi}_o \\%= \begin{bmatrix}2\zeta_1\omega_1 & & \\ & \ddots & \\ & & 2\zeta_K\omega_K\end{bmatrix}\\
       \textbf{k}_o & = \textbf{m}_o^{-1} \boldsymbol{\Phi}_o^T \textbf{K} \boldsymbol{\Phi}_o \\%= \begin{bmatrix}\omega_1^2 & & \\ & \ddots & \\ & & \omega_K^2\end{bmatrix}\\
       \boldsymbol{\theta}_o & = \textbf{m}_o^{-1} \boldsymbol{\Phi}_o^T \boldsymbol{\theta} \\
       \textbf{f}_o & = \textbf{m}_o^{-1} \boldsymbol{\Phi}_o^T \textbf{F}\\
    \end{split}\label{mass_stiff_MOR}
\end{equation}

%%%%%%%%%%%%%%%%%%%%%%%%%%%%%%%%%%%%%%%%%%%%%%%%%%%%%%
%%%%%%%%%%%%%%%%%%%%%% Appendix C %%%%%%%%%%%%%%%%%%%%
%%%%%%%%%%%%%%%%%%%%%%%%%%%%%%%%%%%%%%%%%%%%%%%%%%%%%%
\section{Model Parameters for the Lumped-Element Model}

The parameters for the lumped-element model were derived from the experimental FRFs. The mass of the tip was used to approximate the lumped mass (m). The equivalent stiffness was estimated using the resonant frequency at short circuit (SC) condition ($f_{SC}$) from experiments, and can be computed using the following expression:
\begin{equation}
    k = (2\pi f_{SC})^2m
\end{equation}

The mechanical damping coefficient was determined from the experimental FRFs using the quality factor ($Q$), which is defined as follows:
\begin{equation}
    Q = \frac{f_{SC}}{\Delta f_{SC}}
\end{equation}
where $\Delta f_{SC}$ corresponds to the half-power bandwidth at the SC half-power points. Thus, the damping coefficient is given by the following expresion, 
\begin{equation}
    c=\frac{1}{2Q}\sqrt{mk}
\end{equation}

The electromechanical coupling ($\theta$) is calculated Using the open circuit (OC) resonant frequency $f_{OC}$,
\begin{equation}
    \theta=\sqrt{C_p(m(2\pi f_{OC})^2-k)}
\end{equation}
where $C_p$ is the capacitance of the piezoelectric layers.

\bibliographystyle{model1-num-names}
\bibliography{reference.bib}

\begin{thebibliography}{56}
\expandafter\ifx\csname natexlab\endcsname\relax\def\natexlab#1{#1}\fi
\providecommand{\bibinfo}[2]{#2}
\ifx\xfnm\relax \def\xfnm[#1]{\unskip,\space#1}\fi
%Type = Article
\bibitem[{Lee et~al.(2022)Lee, Lee, Park, Jang, Kim, and Rho}]{lee2022piezoelectric}
\bibinfo{author}{G.~Lee}, \bibinfo{author}{D.~Lee}, \bibinfo{author}{J.~Park}, \bibinfo{author}{Y.~Jang}, \bibinfo{author}{M.~Kim}, \bibinfo{author}{J.~Rho},
\newblock \bibinfo{title}{Piezoelectric energy harvesting using mechanical metamaterials and phononic crystals},
\newblock \bibinfo{journal}{Communications Physics} \bibinfo{volume}{5} (\bibinfo{year}{2022}) \bibinfo{pages}{94}.
%Type = Article
\bibitem[{Han et~al.(2023)Han, Sorokin, Tang, and Cao}]{han2023origami}
\bibinfo{author}{H.~Han}, \bibinfo{author}{V.~Sorokin}, \bibinfo{author}{L.~Tang}, \bibinfo{author}{D.~Cao},
\newblock \bibinfo{title}{Origami-based tunable mechanical memory metamaterial for vibration attenuation},
\newblock \bibinfo{journal}{Mechanical Systems and Signal Processing} \bibinfo{volume}{188} (\bibinfo{year}{2023}) \bibinfo{pages}{110033}.
%Type = Article
\bibitem[{Wei et~al.(2021)Wei, Yang, and Tao}]{wei2021smp}
\bibinfo{author}{Y.-L. Wei}, \bibinfo{author}{Q.-S. Yang}, \bibinfo{author}{R.~Tao},
\newblock \bibinfo{title}{Smp-based chiral auxetic mechanical metamaterial with tunable bandgap function},
\newblock \bibinfo{journal}{International Journal of Mechanical Sciences} \bibinfo{volume}{195} (\bibinfo{year}{2021}) \bibinfo{pages}{106267}.
%Type = Article
\bibitem[{Sugino et~al.(2017)Sugino, Xia, Leadenham, Ruzzene, and Erturk}]{sugino2017general}
\bibinfo{author}{C.~Sugino}, \bibinfo{author}{Y.~Xia}, \bibinfo{author}{S.~Leadenham}, \bibinfo{author}{M.~Ruzzene}, \bibinfo{author}{A.~Erturk},
\newblock \bibinfo{title}{A general theory for bandgap estimation in locally resonant metastructures},
\newblock \bibinfo{journal}{Journal of Sound and Vibration} \bibinfo{volume}{406} (\bibinfo{year}{2017}) \bibinfo{pages}{104--123}.
%Type = Article
\bibitem[{Jansari et~al.(2022)Jansari, Bordas, and Atroshchenko}]{jansari2022design}
\bibinfo{author}{C.~Jansari}, \bibinfo{author}{S.~P. Bordas}, \bibinfo{author}{E.~Atroshchenko},
\newblock \bibinfo{title}{Design of metamaterial-based heat manipulators by isogeometric shape optimization},
\newblock \bibinfo{journal}{International Journal of Heat and Mass Transfer} \bibinfo{volume}{196} (\bibinfo{year}{2022}) \bibinfo{pages}{123201}.
%Type = Article
\bibitem[{Jiao(2020)}]{jiao2020hierarchical}
\bibinfo{author}{P.~Jiao},
\newblock \bibinfo{title}{Hierarchical metastructures with programmable stiffness and zero poisson’s ratio},
\newblock \bibinfo{journal}{APL Materials} \bibinfo{volume}{8} (\bibinfo{year}{2020}).
%Type = Article
\bibitem[{Mizzi et~al.(2018)Mizzi, Mahdi, Titov, Gatt, Attard, Evans, Grima, and Tan}]{mizzi2018mechanical}
\bibinfo{author}{L.~Mizzi}, \bibinfo{author}{E.~Mahdi}, \bibinfo{author}{K.~Titov}, \bibinfo{author}{R.~Gatt}, \bibinfo{author}{D.~Attard}, \bibinfo{author}{K.~E. Evans}, \bibinfo{author}{J.~N. Grima}, \bibinfo{author}{J.-C. Tan},
\newblock \bibinfo{title}{Mechanical metamaterials with star-shaped pores exhibiting negative and zero poisson's ratio},
\newblock \bibinfo{journal}{Materials \& Design} \bibinfo{volume}{146} (\bibinfo{year}{2018}) \bibinfo{pages}{28--37}.
%Type = Article
\bibitem[{Zhang et~al.(2022)Zhang, Bai, and Chen}]{zhang2022dual}
\bibinfo{author}{L.~Zhang}, \bibinfo{author}{Z.~Bai}, \bibinfo{author}{Y.~Chen},
\newblock \bibinfo{title}{Dual-functional hierarchical mechanical metamaterial for vibration insulation and energy absorption},
\newblock \bibinfo{journal}{Engineering Structures} \bibinfo{volume}{271} (\bibinfo{year}{2022}) \bibinfo{pages}{114916}.
%Type = Article
\bibitem[{Xiong et~al.(2023)Xiong, Xu, Wen, Li, and Hosseini}]{xiong2023optimization}
\bibinfo{author}{Y.~Xiong}, \bibinfo{author}{A.~Xu}, \bibinfo{author}{S.~Wen}, \bibinfo{author}{F.~Li}, \bibinfo{author}{S.~M. Hosseini},
\newblock \bibinfo{title}{Optimization of vibration band-gap characteristics of a periodic elastic metamaterial plate},
\newblock \bibinfo{journal}{Mechanics of Advanced Materials and Structures} \bibinfo{volume}{30} (\bibinfo{year}{2023}) \bibinfo{pages}{3204--3214}.
%Type = Article
\bibitem[{Khattak et~al.(2022)Khattak, Sugino, and Erturk}]{khattak2022concurrent}
\bibinfo{author}{M.~M. Khattak}, \bibinfo{author}{C.~Sugino}, \bibinfo{author}{A.~Erturk},
\newblock \bibinfo{title}{Concurrent vibration attenuation and low-power electricity generation in a locally resonant metastructure},
\newblock \bibinfo{journal}{Journal of Intelligent Material Systems and Structures} \bibinfo{volume}{33} (\bibinfo{year}{2022}) \bibinfo{pages}{1990--1999}.
%Type = Article
\bibitem[{Xiao et~al.(2012)Xiao, Wen, and Wen}]{xiao2012broadband}
\bibinfo{author}{Y.~Xiao}, \bibinfo{author}{J.~Wen}, \bibinfo{author}{X.~Wen},
\newblock \bibinfo{title}{Broadband locally resonant beams containing multiple periodic arrays of attached resonators},
\newblock \bibinfo{journal}{Physics Letters A} \bibinfo{volume}{376} (\bibinfo{year}{2012}) \bibinfo{pages}{1384--1390}.
%Type = Article
\bibitem[{Shuguang et~al.(2017)Shuguang, Tianxin, Xudong, and Jialu}]{shuguang2017studies}
\bibinfo{author}{Z.~Shuguang}, \bibinfo{author}{N.~Tianxin}, \bibinfo{author}{W.~Xudong}, \bibinfo{author}{F.~Jialu},
\newblock \bibinfo{title}{Studies of band gaps in flexural vibrations of a locally resonant beam with novel multi-oscillator configuration},
\newblock \bibinfo{journal}{Journal of Vibration and Control} \bibinfo{volume}{23} (\bibinfo{year}{2017}) \bibinfo{pages}{1663--1674}.
%Type = Article
\bibitem[{Sun et~al.(2010)Sun, Du, and Pai}]{sun2010theory}
\bibinfo{author}{H.~Sun}, \bibinfo{author}{X.~Du}, \bibinfo{author}{P.~F. Pai},
\newblock \bibinfo{title}{Theory of metamaterial beams for broadband vibration absorption},
\newblock \bibinfo{journal}{Journal of intelligent material systems and structures} \bibinfo{volume}{21} (\bibinfo{year}{2010}) \bibinfo{pages}{1085--1101}.
%Type = Article
\bibitem[{Sugino et~al.(2016)Sugino, Leadenham, Ruzzene, and Erturk}]{sugino2016mechanism}
\bibinfo{author}{C.~Sugino}, \bibinfo{author}{S.~Leadenham}, \bibinfo{author}{M.~Ruzzene}, \bibinfo{author}{A.~Erturk},
\newblock \bibinfo{title}{On the mechanism of bandgap formation in locally resonant finite elastic metamaterials},
\newblock \bibinfo{journal}{Journal of Applied Physics} \bibinfo{volume}{120} (\bibinfo{year}{2016}).
%Type = Article
\bibitem[{Sachdeva and Ghosh(2024)}]{sachdeva2024aperiodicity}
\bibinfo{author}{R.~Sachdeva}, \bibinfo{author}{D.~Ghosh},
\newblock \bibinfo{title}{Aperiodicity induced robust design of metabeams: Numerical and experimental studies},
\newblock \bibinfo{journal}{International Journal of Mechanical Sciences} \bibinfo{volume}{283} (\bibinfo{year}{2024}) \bibinfo{pages}{109650}.
%Type = Article
\bibitem[{Wang et~al.(2023)Wang, Wan, Hong, Liu, and Li}]{wang2023enhancement}
\bibinfo{author}{G.~Wang}, \bibinfo{author}{S.~Wan}, \bibinfo{author}{J.~Hong}, \bibinfo{author}{S.~Liu}, \bibinfo{author}{X.~Li},
\newblock \bibinfo{title}{Enhancement of the vibration attenuation characteristics in local resonance metamaterial beams: Theory and experiment},
\newblock \bibinfo{journal}{Mechanical Systems and Signal Processing} \bibinfo{volume}{188} (\bibinfo{year}{2023}) \bibinfo{pages}{110036}.
%Type = Article
\bibitem[{Patro et~al.(2023)Patro, Banerjee, and Ramana}]{patro2023vibration}
\bibinfo{author}{S.~R. Patro}, \bibinfo{author}{A.~Banerjee}, \bibinfo{author}{G.~Ramana},
\newblock \bibinfo{title}{Vibration attenuation characteristics of finite locally resonant meta beam: Theory and experiments},
\newblock \bibinfo{journal}{Engineering Structures} \bibinfo{volume}{278} (\bibinfo{year}{2023}) \bibinfo{pages}{115506}.
%Type = Article
\bibitem[{Hu et~al.(2018)Hu, Tang, and Das}]{hu2018internally}
\bibinfo{author}{G.~Hu}, \bibinfo{author}{L.~Tang}, \bibinfo{author}{R.~Das},
\newblock \bibinfo{title}{Internally coupled metamaterial beam for simultaneous vibration suppression and low frequency energy harvesting},
\newblock \bibinfo{journal}{Journal of Applied Physics} \bibinfo{volume}{123} (\bibinfo{year}{2018}).
%Type = Article
\bibitem[{Hu et~al.(2021)Hu, Austin, Sorokin, and Tang}]{hu2021metamaterial}
\bibinfo{author}{G.~Hu}, \bibinfo{author}{A.~C. Austin}, \bibinfo{author}{V.~Sorokin}, \bibinfo{author}{L.~Tang},
\newblock \bibinfo{title}{Metamaterial beam with graded local resonators for broadband vibration suppression},
\newblock \bibinfo{journal}{Mechanical Systems and Signal Processing} \bibinfo{volume}{146} (\bibinfo{year}{2021}) \bibinfo{pages}{106982}.
%Type = Article
\bibitem[{Sugino and Erturk(2018)}]{sugino2018analysis}
\bibinfo{author}{C.~Sugino}, \bibinfo{author}{A.~Erturk},
\newblock \bibinfo{title}{Analysis of multifunctional piezoelectric metastructures for low-frequency bandgap formation and energy harvesting},
\newblock \bibinfo{journal}{Journal of Physics D: Applied Physics} \bibinfo{volume}{51} (\bibinfo{year}{2018}) \bibinfo{pages}{215103}.
%Type = Article
\bibitem[{Sugino et~al.(2018)Sugino, Ruzzene, and Erturk}]{sugino2018merging}
\bibinfo{author}{C.~Sugino}, \bibinfo{author}{M.~Ruzzene}, \bibinfo{author}{A.~Erturk},
\newblock \bibinfo{title}{Merging mechanical and electromechanical bandgaps in locally resonant metamaterials and metastructures},
\newblock \bibinfo{journal}{Journal of the Mechanics and Physics of Solids} \bibinfo{volume}{116} (\bibinfo{year}{2018}) \bibinfo{pages}{323--333}.
%Type = Article
\bibitem[{Zhao et~al.(2022)Zhao, Thomsen, De~Ponti, Riva, Van~Damme, Bergamini, Chatzi, and Colombi}]{zhao2022graded}
\bibinfo{author}{B.~Zhao}, \bibinfo{author}{H.~R. Thomsen}, \bibinfo{author}{J.~M. De~Ponti}, \bibinfo{author}{E.~Riva}, \bibinfo{author}{B.~Van~Damme}, \bibinfo{author}{A.~Bergamini}, \bibinfo{author}{E.~Chatzi}, \bibinfo{author}{A.~Colombi},
\newblock \bibinfo{title}{A graded metamaterial for broadband and high-capability piezoelectric energy harvesting},
\newblock \bibinfo{journal}{Energy Conversion and Management} \bibinfo{volume}{269} (\bibinfo{year}{2022}) \bibinfo{pages}{116056}.
%Type = Article
\bibitem[{Covaci and Gontean(2020)}]{covaci2020piezoelectric}
\bibinfo{author}{C.~Covaci}, \bibinfo{author}{A.~Gontean},
\newblock \bibinfo{title}{Piezoelectric energy harvesting solutions: A review},
\newblock \bibinfo{journal}{Sensors} \bibinfo{volume}{20} (\bibinfo{year}{2020}) \bibinfo{pages}{3512}.
%Type = Article
\bibitem[{Sezer and Ko{\c{c}}(2021)}]{sezer2021comprehensive}
\bibinfo{author}{N.~Sezer}, \bibinfo{author}{M.~Ko{\c{c}}},
\newblock \bibinfo{title}{A comprehensive review on the state-of-the-art of piezoelectric energy harvesting},
\newblock \bibinfo{journal}{Nano Energy} \bibinfo{volume}{80} (\bibinfo{year}{2021}) \bibinfo{pages}{105567}.
%Type = Article
\bibitem[{Izadgoshasb(2021)}]{izadgoshasb2021piezoelectric}
\bibinfo{author}{I.~Izadgoshasb},
\newblock \bibinfo{title}{Piezoelectric energy harvesting towards self-powered internet of things (iot) sensors in smart cities},
\newblock \bibinfo{journal}{Sensors} \bibinfo{volume}{21} (\bibinfo{year}{2021}) \bibinfo{pages}{8332}.
%Type = Book
\bibitem[{Erturk and Inman(2011)}]{erturk2011piezoelectric}
\bibinfo{author}{A.~Erturk}, \bibinfo{author}{D.~J. Inman}, \bibinfo{title}{Piezoelectric energy harvesting}, \bibinfo{publisher}{John Wiley \& Sons}, \bibinfo{year}{2011}.
%Type = Article
\bibitem[{Erturk and Inman(2008)}]{erturk2008distributed}
\bibinfo{author}{A.~Erturk}, \bibinfo{author}{D.~J. Inman},
\newblock \bibinfo{title}{A distributed parameter electromechanical model for cantilevered piezoelectric energy harvesters},
\newblock \bibinfo{journal}{Journal of vibration and acoustics} \bibinfo{volume}{130} (\bibinfo{year}{2008}) \bibinfo{pages}{041002}.
%Type = Article
\bibitem[{El-Borgi et~al.(2020)El-Borgi, Fernandes, Rajendran, Yazbeck, Boyd, and Lagoudas}]{el2020multiple}
\bibinfo{author}{S.~El-Borgi}, \bibinfo{author}{R.~Fernandes}, \bibinfo{author}{P.~Rajendran}, \bibinfo{author}{R.~Yazbeck}, \bibinfo{author}{J.~Boyd}, \bibinfo{author}{D.~Lagoudas},
\newblock \bibinfo{title}{Multiple bandgap formation in a locally resonant linear metamaterial beam: Theory and experiments},
\newblock \bibinfo{journal}{Journal of Sound and Vibration} \bibinfo{volume}{488} (\bibinfo{year}{2020}) \bibinfo{pages}{115647}.
%Type = Article
\bibitem[{Hu et~al.(2017)Hu, Tang, Banerjee, and Das}]{hu2017metastructure}
\bibinfo{author}{G.~Hu}, \bibinfo{author}{L.~Tang}, \bibinfo{author}{A.~Banerjee}, \bibinfo{author}{R.~Das},
\newblock \bibinfo{title}{Metastructure with piezoelectric element for simultaneous vibration suppression and energy harvesting},
\newblock \bibinfo{journal}{Journal of Vibration and Acoustics} \bibinfo{volume}{139} (\bibinfo{year}{2017}) \bibinfo{pages}{011012}.
%Type = Article
\bibitem[{Zhang et~al.(2024)Zhang, Chen, Karimi, Li, Saydam, and Hassan}]{zhang2024numerical}
\bibinfo{author}{H.~Zhang}, \bibinfo{author}{S.~Chen}, \bibinfo{author}{M.~Karimi}, \bibinfo{author}{B.~Li}, \bibinfo{author}{S.~Saydam}, \bibinfo{author}{M.~Hassan},
\newblock \bibinfo{title}{Numerical and experimental investigation of an auxetic piezoelectric energy harvester with frequency self-tuning capability},
\newblock \bibinfo{journal}{Smart Materials and Structures} \bibinfo{volume}{33} (\bibinfo{year}{2024}) \bibinfo{pages}{055022}.
%Type = Article
\bibitem[{Aghakhani et~al.(2020)Aghakhani, Gozum, and Basdogan}]{aghakhani2020modal}
\bibinfo{author}{A.~Aghakhani}, \bibinfo{author}{M.~M. Gozum}, \bibinfo{author}{I.~Basdogan},
\newblock \bibinfo{title}{Modal analysis of finite-size piezoelectric metamaterial plates},
\newblock \bibinfo{journal}{Journal of Physics D: Applied Physics} \bibinfo{volume}{53} (\bibinfo{year}{2020}) \bibinfo{pages}{505304}.
%Type = Article
\bibitem[{Dwivedi et~al.(2021)Dwivedi, Banerjee, Adhikari, and Bhattacharya}]{dwivedi2021optimal}
\bibinfo{author}{A.~Dwivedi}, \bibinfo{author}{A.~Banerjee}, \bibinfo{author}{S.~Adhikari}, \bibinfo{author}{B.~Bhattacharya},
\newblock \bibinfo{title}{Optimal electromechanical bandgaps in piezo-embedded mechanical metamaterials},
\newblock \bibinfo{journal}{International Journal of Mechanics and Materials in Design} \bibinfo{volume}{17} (\bibinfo{year}{2021}) \bibinfo{pages}{419--439}.
%Type = Article
\bibitem[{Mao et~al.(2024)Mao, Gao, Zhu, Gao, and Qu}]{mao2024analytical}
\bibinfo{author}{J.~Mao}, \bibinfo{author}{H.~Gao}, \bibinfo{author}{J.~Zhu}, \bibinfo{author}{P.~Gao}, \bibinfo{author}{Y.~Qu},
\newblock \bibinfo{title}{Analytical modeling of piezoelectric meta-beams with unidirectional circuit for broadband vibration attenuation},
\newblock \bibinfo{journal}{Applied Mathematics and Mechanics} \bibinfo{volume}{45} (\bibinfo{year}{2024}) \bibinfo{pages}{1665--1684}.
%Type = Article
\bibitem[{Jian et~al.(2022)Jian, Tang, Hu, Li, and Aw}]{jian2022design}
\bibinfo{author}{Y.~Jian}, \bibinfo{author}{L.~Tang}, \bibinfo{author}{G.~Hu}, \bibinfo{author}{Z.~Li}, \bibinfo{author}{K.~C. Aw},
\newblock \bibinfo{title}{Design of graded piezoelectric metamaterial beam with spatial variation of electrodes},
\newblock \bibinfo{journal}{International Journal of Mechanical Sciences} \bibinfo{volume}{218} (\bibinfo{year}{2022}) \bibinfo{pages}{107068}.
%Type = Article
\bibitem[{Alshaqaq and Erturk(2020)}]{alshaqaq2020graded}
\bibinfo{author}{M.~Alshaqaq}, \bibinfo{author}{A.~Erturk},
\newblock \bibinfo{title}{Graded multifunctional piezoelectric metastructures for wideband vibration attenuation and energy harvesting},
\newblock \bibinfo{journal}{Smart Materials and Structures} \bibinfo{volume}{30} (\bibinfo{year}{2020}) \bibinfo{pages}{015029}.
%Type = Article
\bibitem[{Liu et~al.(2023)Liu, Han, and Liu}]{liu2023broadband}
\bibinfo{author}{Y.~Liu}, \bibinfo{author}{C.~Han}, \bibinfo{author}{D.~Liu},
\newblock \bibinfo{title}{Broadband vibration suppression of graded/disorder piezoelectric metamaterials},
\newblock \bibinfo{journal}{Mechanics of Advanced Materials and Structures} \bibinfo{volume}{30} (\bibinfo{year}{2023}) \bibinfo{pages}{710--723}.
%Type = Article
\bibitem[{Jian et~al.(2023)Jian, Hu, Tang, Tang, Abdi, and Aw}]{jian2023analytical}
\bibinfo{author}{Y.~Jian}, \bibinfo{author}{G.~Hu}, \bibinfo{author}{L.~Tang}, \bibinfo{author}{W.~Tang}, \bibinfo{author}{M.~Abdi}, \bibinfo{author}{K.~C. Aw},
\newblock \bibinfo{title}{Analytical and experimental study of a metamaterial beam with grading piezoelectric transducers for vibration attenuation band widening},
\newblock \bibinfo{journal}{Engineering Structures} \bibinfo{volume}{275} (\bibinfo{year}{2023}) \bibinfo{pages}{115091}.
%Type = Article
\bibitem[{Schimidt et~al.(2023)Schimidt, de~Sousa, and De~Marqui~Junior}]{schimidt2023piezoelectric}
\bibinfo{author}{C.~S. Schimidt}, \bibinfo{author}{V.~C. de~Sousa}, \bibinfo{author}{C.~De~Marqui~Junior},
\newblock \bibinfo{title}{Piezoelectric energy harvesting in graded elastic metastructures using continuous and segmented electrodes},
\newblock \bibinfo{journal}{Journal of the Brazilian Society of Mechanical Sciences and Engineering} \bibinfo{volume}{45} (\bibinfo{year}{2023}) \bibinfo{pages}{329}.
%Type = Article
\bibitem[{Jian et~al.(2022)Jian, Tang, Hu, Wang, and Aw}]{jian2022adaptive}
\bibinfo{author}{Y.~Jian}, \bibinfo{author}{L.~Tang}, \bibinfo{author}{G.~Hu}, \bibinfo{author}{Y.~Wang}, \bibinfo{author}{K.~C. Aw},
\newblock \bibinfo{title}{Adaptive genetic algorithm enabled tailoring of piezoelectric metamaterials for optimal vibration attenuation},
\newblock \bibinfo{journal}{Smart Materials and Structures} \bibinfo{volume}{31} (\bibinfo{year}{2022}) \bibinfo{pages}{075026}.
%Type = Article
\bibitem[{Junior et~al.(2009)Junior, Erturk, and Inman}]{junior2009electromechanical}
\bibinfo{author}{C.~D.~M. Junior}, \bibinfo{author}{A.~Erturk}, \bibinfo{author}{D.~J. Inman},
\newblock \bibinfo{title}{An electromechanical finite element model for piezoelectric energy harvester plates},
\newblock \bibinfo{journal}{Journal of Sound and Vibration} \bibinfo{volume}{327} (\bibinfo{year}{2009}) \bibinfo{pages}{9--25}.
%Type = Article
\bibitem[{Peralta et~al.(2020)Peralta, Ruiz, Natarajan, and Atroshchenko}]{peralta2020parametric}
\bibinfo{author}{P.~Peralta}, \bibinfo{author}{R.~Ruiz}, \bibinfo{author}{S.~Natarajan}, \bibinfo{author}{E.~Atroshchenko},
\newblock \bibinfo{title}{Parametric study and shape optimization of piezoelectric energy harvesters by isogeometric analysis and kriging metamodeling},
\newblock \bibinfo{journal}{Journal of Sound and Vibration} \bibinfo{volume}{484} (\bibinfo{year}{2020}) \bibinfo{pages}{115521}.
%Type = Article
\bibitem[{Nguyen et~al.(2017)Nguyen, Atroshchenko, Nguyen-Xuan, and Vo}]{nguyen2017geometrically}
\bibinfo{author}{H.~X. Nguyen}, \bibinfo{author}{E.~Atroshchenko}, \bibinfo{author}{H.~Nguyen-Xuan}, \bibinfo{author}{T.~P. Vo},
\newblock \bibinfo{title}{Geometrically nonlinear isogeometric analysis of functionally graded microplates with the modified couple stress theory},
\newblock \bibinfo{journal}{Computers \& Structures} \bibinfo{volume}{193} (\bibinfo{year}{2017}) \bibinfo{pages}{110--127}.
%Type = Article
\bibitem[{Hughes et~al.(2005)Hughes, Cottrell, and Bazilevs}]{hughes2005isogeometric}
\bibinfo{author}{T.~J. Hughes}, \bibinfo{author}{J.~A. Cottrell}, \bibinfo{author}{Y.~Bazilevs},
\newblock \bibinfo{title}{Isogeometric analysis: Cad, finite elements, nurbs, exact geometry and mesh refinement},
\newblock \bibinfo{journal}{Computer methods in applied mechanics and engineering} \bibinfo{volume}{194} (\bibinfo{year}{2005}) \bibinfo{pages}{4135--4195}.
%Type = Article
\bibitem[{Schu{\ss} et~al.(2019)Schu{\ss}, Dittmann, Wohlmuth, Klinkel, and Hesch}]{schuss2019multi}
\bibinfo{author}{S.~Schu{\ss}}, \bibinfo{author}{M.~Dittmann}, \bibinfo{author}{B.~Wohlmuth}, \bibinfo{author}{S.~Klinkel}, \bibinfo{author}{C.~Hesch},
\newblock \bibinfo{title}{Multi-patch isogeometric analysis for kirchhoff--love shell elements},
\newblock \bibinfo{journal}{Computer Methods in Applied Mechanics and Engineering} \bibinfo{volume}{349} (\bibinfo{year}{2019}) \bibinfo{pages}{91--116}.
%Type = Article
\bibitem[{Guo and Ruess(2015)}]{guo2015nitsche}
\bibinfo{author}{Y.~Guo}, \bibinfo{author}{M.~Ruess},
\newblock \bibinfo{title}{Nitsche’s method for a coupling of isogeometric thin shells and blended shell structures},
\newblock \bibinfo{journal}{Computer Methods in Applied Mechanics and Engineering} \bibinfo{volume}{284} (\bibinfo{year}{2015}) \bibinfo{pages}{881--905}.
%Type = Article
\bibitem[{Benzaken et~al.(2021)Benzaken, Evans, McCormick, and Tamstorf}]{benzaken2021nitsche}
\bibinfo{author}{J.~Benzaken}, \bibinfo{author}{J.~A. Evans}, \bibinfo{author}{S.~F. McCormick}, \bibinfo{author}{R.~Tamstorf},
\newblock \bibinfo{title}{Nitsche’s method for linear kirchhoff--love shells: Formulation, error analysis, and verification},
\newblock \bibinfo{journal}{Computer Methods in Applied Mechanics and Engineering} \bibinfo{volume}{374} (\bibinfo{year}{2021}) \bibinfo{pages}{113544}.
%Type = Article
\bibitem[{Guo and Ruess(2015)}]{guo2015weak}
\bibinfo{author}{Y.~Guo}, \bibinfo{author}{M.~Ruess},
\newblock \bibinfo{title}{Weak dirichlet boundary conditions for trimmed thin isogeometric shells},
\newblock \bibinfo{journal}{Computers \& Mathematics with Applications} \bibinfo{volume}{70} (\bibinfo{year}{2015}) \bibinfo{pages}{1425--1440}.
%Type = Book
\bibitem[{Crandall(1968)}]{crandall1968dynamics}
\bibinfo{author}{S.~H. Crandall}, \bibinfo{title}{Dynamics of mechanical and electromechanical systems}, \bibinfo{publisher}{McGraw-Hill}, \bibinfo{year}{1968}.
%Type = Book
\bibitem[{Bittencourt(2014)}]{bittencourt2014computational}
\bibinfo{author}{M.~L. Bittencourt}, \bibinfo{title}{Computational solid mechanics: Variational formulation and high order approximation}, \bibinfo{publisher}{CRC Press}, \bibinfo{year}{2014}.
%Type = Article
\bibitem[{Guo(2016)}]{guo2016isogeometric}
\bibinfo{author}{Y.~Guo},
\newblock \bibinfo{title}{Isogeometric analysis for thin-walled composite structures}  (\bibinfo{year}{2016}).
%Type = Book
\bibitem[{Reddy(2006)}]{reddy2006theory}
\bibinfo{author}{J.~N. Reddy}, \bibinfo{title}{Theory and analysis of elastic plates and shells}, \bibinfo{publisher}{CRC press}, \bibinfo{year}{2006}.
%Type = Article
\bibitem[{Erturk and Inman(2009)}]{erturk2009experimentally}
\bibinfo{author}{A.~Erturk}, \bibinfo{author}{D.~J. Inman},
\newblock \bibinfo{title}{An experimentally validated bimorph cantilever model for piezoelectric energy harvesting from base excitations},
\newblock \bibinfo{journal}{Smart materials and structures} \bibinfo{volume}{18} (\bibinfo{year}{2009}) \bibinfo{pages}{025009}.
%Type = Article
\bibitem[{Peralta-Braz et~al.(2023)Peralta-Braz, Alamdari, Ruiz, Atroshchenko, and Hassan}]{peralta2023design}
\bibinfo{author}{P.~Peralta-Braz}, \bibinfo{author}{M.~M. Alamdari}, \bibinfo{author}{R.~O. Ruiz}, \bibinfo{author}{E.~Atroshchenko}, \bibinfo{author}{M.~Hassan},
\newblock \bibinfo{title}{Design optimisation of piezoelectric energy harvesters for bridge infrastructure},
\newblock \bibinfo{journal}{Mechanical Systems and Signal Processing} \bibinfo{volume}{205} (\bibinfo{year}{2023}) \bibinfo{pages}{110823}.
%Type = Article
\bibitem[{Rahimzadeh et~al.(2021)Rahimzadeh, Samadi, and Mohammadi}]{rahimzadeh2021analysis}
\bibinfo{author}{M.~Rahimzadeh}, \bibinfo{author}{H.~Samadi}, \bibinfo{author}{N.~S. Mohammadi},
\newblock \bibinfo{title}{Analysis of energy harvesting enhancement in piezoelectric unimorph cantilevers},
\newblock \bibinfo{journal}{Sensors} \bibinfo{volume}{21} (\bibinfo{year}{2021}) \bibinfo{pages}{8463}.
%Type = Article
\bibitem[{Xu et~al.(2022)Xu, Gao, Li, and Jin}]{xu2022design}
\bibinfo{author}{Q.~Xu}, \bibinfo{author}{A.~Gao}, \bibinfo{author}{Y.~Li}, \bibinfo{author}{Y.~Jin},
\newblock \bibinfo{title}{Design and optimization of piezoelectric cantilever beam vibration energy harvester},
\newblock \bibinfo{journal}{Micromachines} \bibinfo{volume}{13} (\bibinfo{year}{2022}) \bibinfo{pages}{675}.
%Type = Article
\bibitem[{Sugino et~al.(2020)Sugino, Ruzzene, and Erturk}]{sugino2020analytical}
\bibinfo{author}{C.~Sugino}, \bibinfo{author}{M.~Ruzzene}, \bibinfo{author}{A.~Erturk},
\newblock \bibinfo{title}{An analytical framework for locally resonant piezoelectric metamaterial plates},
\newblock \bibinfo{journal}{International Journal of Solids and Structures} \bibinfo{volume}{182} (\bibinfo{year}{2020}) \bibinfo{pages}{281--294}.

\end{thebibliography}

\end{document}